\documentclass{aa}  

\usepackage{graphicx}
\usepackage{txfonts}
\usepackage[breaklinks,colorlinks,pdfa=true]{hyperref}

\newcommand{\var}[1]{{\textrm{Var}\;\!(#1)}}
\newcommand{\cov}[1]{{\textrm{Cov}\;\!(#1)}}
\newcommand{\ev}[1]{\langle{#1}\rangle}

\newcommand{\nside}{\ensuremath N_{\text{side}}}

\begin{document}
   \title{Astrophysical foreground cleanup using non-local means}
   
   \author{Guillermo F.\ Quispe Pe\~na
      \inst{1}\fnmsep\thanks{\email{gfq@sfu.ca}}
   \and
      Andrei V.\ Frolov
      \inst{1}\fnmsep\thanks{\email{frolov@sfu.ca}}
   }
   
   \institute{Department of Physics, Simon Fraser University,
      8888 University Drive, Burnaby BC V5A 1S6, Canada
   }

   \date{Received May 31, 2023; accepted ???}

 
  \abstract
   {To create high-fidelity cosmic microwave background maps, current component separation methods rely on availability of information on different foreground components, usually through multi-band frequency coverage of the instrument. Internal linear combination (ILC) methods provide an unbiased estimators for CMB which are easy to implement, but component separation quality crucially depends on the signal to noise ratio of the input maps. In the present paper, we describe a non-linear filter which significantly improves signal to noise ratio for astrophysical foreground maps, while having minimal signal attenuation.}
   {We develop an efficient non-linear filter along the lines of non-local means used in digital imaging research which is suitable (and fast enough) for application to full resolution Planck foreground maps, and evaluate it performance in map and spectral domains.}
   {Noise reduction is achieved by averaging ``similar'' pixels in the map. We construct the rotationally-invariant feature vector space and compute the similarity metric on it for the case of non-Gaussian signal contaminated by an additive Gaussian noise.}
   {The proposed filter has two tuneable parameters, and with minimal tweaking achieves a factor of two improvement in signal to noise spectral density in Planck dust maps. A particularly desirable feature is that signal loss is extremely small at all scales.}
   {}
   

   \maketitle
%

\section{Introduction}
\label{sec:intro}

The Cosmic Microwave Background (CMB) provides essential information about all the epochs of our Universe and plays a fundamental role in understanding its structure and dynamical evolution, e.g.\ see recent overview by \cite{2020A&A...641A...1P}. Precise measurements and theoretical predictions enable an accurate reconstruction of CMB sky maps that aid in measuring cosmological parameters \citep{1994PhRvL..72...13B, 2020A&A...641A...6P} and constraining various physical phenomena \citep{2020A&A...641A...1P}. The mapping of small anisotropies in intensity and polarization of the CMB has had the most significant impact, providing stringent constraints on models of the early Universe \citep{2013ApJS..208...19H, 2020A&A...641A..10P, 2020A&A...641A...9P, 2020A&A...641A...7P}.

In recent decades, an important objective of CMB experiments has been the measurement of CMB polarization, with a particular focus on detecting curl modes known as $B$-modes, e.g.\ as discussed by \cite{2023arXiv230204276W}. The detection of these modes would carry significant implications, as they could potentially provide evidence of primordial gravitational waves and enhance our understanding of the early Universe \citep{1993PhRvL..71..324C}. However, observing the CMB is challenging due to the presence of local contamination from various astrophysical sources, collectively referred to as CMB foregrounds. Some of these foreground emissions exhibit polarization, including $B$-modes, which introduce contamination into our observations of the primary CMB $B$-modes \citep{2020A&A...641A..11P, 2021PhRvL.127o1301A}. Consequently, a crucial step in analyzing CMB data involves effectively separating the (polarized) foreground emissions from the overall observed sky signal in order to retrieve valuable information from the CMB \citep{2008A&A...491..597L, 2020A&A...641A...4P}. This could be accomplished either on the map or anisotropy spectrum levels of data reduction.

The characterization of astrophysical components is currently based on their frequency dependence, which enables us to effectively separate them and obtain clean maps of the CMB. Recent advancements in sensitivity and frequency coverage of CMB experiments have led to significant progress in component separation techniques, which can be broadly categorized into maximal likelihood estimators, usually Gibbs samplers \citep{2004PhRvD..70h3511W, 2008ApJ...676...10E} and unbiased linear estimators, usually referred to as internal linear combination (ILC) in the CMB literature \citep{2003MNRAS.345.1101M, 2003MNRAS.346.1089D, 2008arXiv0803.1814C, 2011MNRAS.418..467R}. However, even the most sophisticated foreground removal processes cannot completely eliminate instrumental noise and residual foreground contamination from the final data. Accurate frequency modelling of astrophysical foregrounds is essential for maximal likelihood estimators, whereas ILC estimators could potentially be biased by the noise. As any inaccuracies may introduce biases in the estimation of cosmological parameters, it is crucial to continually improve the characterization of foreground components and enhance their signal-to-noise ratios to ensure reliable and accurate CMB analysis. In this paper, we present a new method which significantly attenuates the noise while keeping the signal mostly unaffected for strongly non-Gaussian data.

This paper is structured as follows. In Section \ref{sec:nlmean}, we introduce a denoising algorithm known as non-local means, initially proposed by \citet{Buades:2005}. We extend this filter by incorporating covariant functions that capture morphological features of the input map on a sphere, thereby modifying the non-local averaging procedure. The specific set of functions and the criteria for their selection are outlined in Section \ref{sec:feature}. The application of our filter to Planck thermal dust and CMB maps is presented in Sections \ref{sec:dust} and \ref{sec:cmb}, where we discuss the obtained results. Possible extensions to polarization data are considered in Section \ref{sec:polar}. Finally, in Section \ref{sec:conclusion}, we discuss our results. Technical details concerning calculation of the covariance matrix associated with our chosen feature space estimators can be found in Appendix \ref{sec:app:variance}. Appendix \ref{sec:app:correlation} contains derivation of the characteristic parameters for the angular two-point correlation functions of Gaussian noise models employed in this study. Appendix \ref{sec:app:dust} describes preprocessing of the $353$GHz maps that we used as test samples.

\section{Non-local means}
\label{sec:nlmean}

Sky emission maps are essentially digital images that can be represented as arrays of real numbers. Each pixel in such an image can be expressed as a pair $(i, d_{i})$, where $i$ denotes a point on a 2-dimensional grid and $d_i$ represents the associated real value. Common pixelization scheme of a sphere which is used for most CMB data is HEALPix (Hierarchical Equal Area isoLatitude Pixelization) by \cite{2005ApJ...622..759G}. The accuracy of digital images is often limited by the presence of noise. In the context of measured sky emission data, this noise can arise from various sources such as photon noise, phonon noise, and glitch residuals. To model the observed data $d$, we can express it as the sum of the true signal value $s$ and the noise perturbation $n$, yielding the equation
\begin{equation}
  d = s + n .
\end{equation}
While noise is typically dependent on instrument properties and scan strategy, a commonly employed approximation in data analysis is to assume a zero-mean additive Gaussian noise model with known covariance. In the simplest and most common case, it is diagonal in pixel space, i.e.\ the pixel noise is independent, although the variance can vary from pixel to pixel. Signal could either be a Gaussian random field, such as the case for CMB, or completely non-Gaussian and full of features, as most astrophysical foregrounds are.

In order to denoise the image and restore the underlying true signal $s$, we employ the non-local means denoising method by \cite{Buades:2005}. This method operates on the noisy image $d$ and estimates the true value $s_i$ at each pixel $i$ by computing the mean of the values of all pixels whose neighbourhood exhibits \textit{similarity} to the neighbourhood of pixel $i$. In contrast with the usual smoothing, where the averaging weight depends on pixel \textit{proximity}, the assessment of similarity between different pixels in the image is performed by comparing their values, as well as other characteristics of the pixel neighbourhood, the effective size of which can be controlled by the Gaussian smoothing beam
\begin{equation}
  \tilde{s} = b*d,
\end{equation}
where the convolution operation $*$ is performed using a Gaussian convolution kernel $b$. The width of the kernel depends on a free smoothing parameter, often specified as the full width at half maximum (FWHM), allowing for control over the level of smoothing applied to the image for feature identification. In the simplest version, the estimated value $s_i$ is given by
\begin{equation}\label{2.3}
    s_{i}=\frac{\sum_{j}w(\tilde{s}_{i},\tilde{s}_{j})\,d_{j}}{\sum_{j}w(\tilde{s}_{i},\tilde{s}_{j})} ,
\end{equation}
where the sum is over the entire map, and the weight
\begin{equation}\label{2.4}
    w(\tilde{s}_{i},\tilde{s}_{j})=\exp\left[-\frac{1}{2}\frac{(\tilde{s}_{i}-\tilde{s}_{j})^{2}}{h^{2}}\right]
\end{equation}
quantifies the similarity between pixels $i$ and $j$ by comparing their corresponding values in the Gaussian-smoothed map $\tilde{s}$. The parameter $h$ determines the width of the Gaussian kernel in the feature space for filtering the map, and could be defined as
\begin{equation}
    h^{2}=\alpha^{2}\var{\delta\tilde{s}},
\end{equation}
where $\alpha$ is a user-specified parameter that determines the filtering strength, and $\delta\tilde{s}$ is the noise component of the smoothed map. As the noise contribution might not be known outright, it could be useful to bootstrap it from the map itself as $d-\tilde{s}$, and tweak parameter $\alpha$ for desired filter strength.

In contrast to just using a smoothed pixel value, we propose a refined method to compute the similarity between Gaussian neighbourhoods by incorporating additional morphological features extracted from the input map $d$. This is achieved by constructing a feature space, represented by a collection of maps ${{\cal F}^{(n)}}$ that capture relevant morphological information from $d$. In the original implementation of \cite{Buades:2005}, field values in a square pixel neighbourhood were used as a feature vector. This is obviously not optimal for statistically isotropic maps such as CMB, since it is not rotationally invariant, and in any case problematic with HEALPix as the number of nearest neighbours of a pixel varies. Instead, we propose to use covariant invariants of the map and its derivatives, which could be graded on field and derivative power. These maps are combined to form a feature vector field ${\cal F}=[{\cal F}^{(1)},{\cal F}^{(2)},\cdots]^{T}$. The expansion of the feature space enables the incorporation of additional information beyond the comparison of values in $\tilde{s}$ alone. Consequently, the estimation of the true value $s_i$ in Eq. \eqref{2.3} can be generalized as
\begin{equation} \label{eq:nlmean}
  s_i = \frac{\sum_j w({\cal F}_i,{\cal F}_j)\, d_j}{\sum_j w({\cal F}_i,{\cal F}_j)} ,
\end{equation}
where the weight function now evaluates the similarity between pixels $i$ and $j$ based on the comparison of their corresponding feature vectors in ${\cal F}$. The weight function in Eq. \eqref{2.4} can be generalized as
\begin{equation}\label{eq:weight}
  w({\cal F}_i,{\cal F}_j) = \exp\left[-\frac{1}{2} ({\cal F}_i-{\cal F}_j)^T \omega^{-2} \, ({\cal F}_i-{\cal F}_j)\right] ,
\end{equation}
where $\omega^{-2}$ defines a similarity metric on the feature space with
\begin{equation}
   \omega^{2} = \alpha^{2}\, \var{\delta{\cal F}} .
\end{equation}
The feature space extends the 1-dimensional variance $\var{\delta\tilde{s}}$ to a multidimensional covariance matrix of noise perturbations $\var{\delta\cal F}$, capturing the statistical relationships among the feature estimators. Once again, the adjustable parameter $\alpha$ controls the degree of filtering. Further details regarding the calculation of the covariance matrix for the specific feature space employed in this study can be found in Appendix \ref{sec:app:variance}.

As the sum (\ref{eq:nlmean}) has to be done for every pixel, computational cost of the algorithm scales as a number of pixels squared, and increases with dimensionality of the feature space. While expensive, the algorithm is trivially paralleizable, and lends itself well for computation on GPUs. To increase computational speed and enforce some locality, the sum (\ref{eq:nlmean}) could be restricted to a specific neighbourhood of a pixel (say, within certain angular radius), or even outfitted with a weight based on proximity, providing a bridge to the usual convolutions. In essence, the generalized non-local means (\ref{eq:nlmean}) is an extension of a regular convolution to a distance measured on a surface embedded into a higher-dimensional space-feature manifold.

\section{Feature space}
\label{sec:feature}

The feature space ${{\cal F}^{(n)}}$ should include information that captures relevant and non-redundant morphological information of the maps, in particular hot and cold spots in emission. This inclusion aims to enhance the accuracy of pixel similarity comparisons. In the previous section, we discussed limitations of the original non-local means algorithm, where the feature vector is tied down to a particular pixelization scheme. This does not make sense for our application, and instead a tower of differential invariants seems like a natural choice. Starting from the scalar field $\varphi$ and grading by the number of derivatives applied, these would be $\varphi$, $(\nabla \varphi)^2$, $\Delta\varphi$, $\varphi_{;ab} \varphi^{;a} \varphi^{;b}$, $\varphi_{;ab} \varphi^{;a} \epsilon^{bc}\varphi_{;c}$, $\varphi_{;ab} \epsilon^{ac} \varphi_{;c} \epsilon^{bd}\varphi_{;d}$, $\varphi_{;ab}\varphi^{;ab}$ and so on. Here semicolon denotes covariant derivative $\nabla$ on a sphere, $\Delta\equiv\nabla_a\nabla^a$ is Laplace operator, while $\epsilon_{ab}$ is a total antisymmetric symbol in two dimensions, used to make duals. Not all of these invariants are of degree one with respect to the field $\varphi$, but they could be made so by taking fractional power or by dividing by lower-order invariants. The number of combinations rapidly increases with the rank of derivative operator. 

The trick is to pick as few as possible, while still having access to enough morphological information. Besides the field value $\varphi$, the obvious candidate linear in $\varphi$ is $|\nabla\varphi| = \sqrt{\varphi_{;a}\varphi^{;a}}$, vanishing of which distinguishes peaks. The next in line are three expressions with four derivatives normalized to be linear in $\varphi$ by dividing them by $(\nabla\varphi)^2$, namely $\varphi_{;ab} \varphi^{;a} \varphi^{;b}/(\nabla\varphi)^2$, $\varphi_{;ab} \varphi^{;a} \epsilon^{bc}\varphi_{;c}/(\nabla\varphi)^2$, and $\varphi_{;ab} \epsilon^{ac} \varphi_{;c} \epsilon^{bd}\varphi_{;d}/(\nabla\varphi)^2$. These have the meaning of components of the field Hessian matrix written in orthonormal basis $\vec{g}^{(1)}_a = \varphi_{;a}/|\nabla\varphi|$ and $\vec{g}^{(2)}_a = \epsilon_{ab}\varphi^{;b}/|\nabla\varphi|$ aligned with the field gradient direction. Notably, the third expression enters Minkowski functional integral ${\cal I}_2$ in \cite{1998MNRAS.297..355S}, while zero set of the second one corresponds to the map skeleton, as described in \cite{2006MNRAS.366.1201N}, and is most interesting morphologically. After some experimentation, we settled on the feature space consisting of the field value, length of its gradient, and the skeleton invariant. As we mentioned in the previous section, these are constructed from the smoothed field $\tilde{s}$.

The selection of these features was guided in part by the objective of minimizing the complexity of the covariance matrix $\var{\delta\cal F}$ to reduce computational cost, while still incorporating useful information provided by the Gaussian-smoothed map $\tilde{s}$ and its first and second covariant derivatives. The covariant derivatives at a point $(\theta,\phi)$ on the unit sphere, projected onto orthonormal basis vectors $\vec{e}^{(1)} = \hat{\theta}$ and $\vec{e}^{(2)} = \hat{\phi}$, are
\begin{equation}
\begin{aligned}
    \tilde{s}_{;1}&=\frac{\partial\tilde{s}}{\partial\theta} , \\
    \tilde{s}_{;2}&=\frac{1}{\sin\theta}\frac{\partial\tilde{s}}{\partial\phi} , \\
    \tilde{s}_{;11}&=\frac{\partial^{2}\tilde{s}}{\partial\theta^{2}} , \\
    \tilde{s}_{;12}&=\frac{1}{\sin\theta}\frac{\partial^{2}\tilde{s}}{\partial\theta\partial\phi}-\frac{\cos\theta}{\sin^{2}\theta}\frac{\partial\tilde{s}}{\partial\phi} , \\
    \tilde{s}_{;22}&=\frac{1}{\sin^{2}\theta}\frac{\partial^{2}\tilde{s}}{\partial\phi^{2}}+\frac{\cos\theta}{\sin\theta}\frac{\partial\tilde{s}}{\partial\theta} ,
    \end{aligned}
\end{equation}
and can be numerically evaluated using standard HEALPix routines. Thus the first and second features can be computed as 
\begin{equation} \label{eq:feature:12}
  {\cal F}^{(1)} = \tilde{s}, \hspace{1em}
  {\cal F}^{(2)} = |\nabla\tilde{s}| = \sqrt{\tilde{s}_{;1}^2 + \tilde{s}_{;2}^2} .
\end{equation}
As for the third feature, it is obtained from the second covariant derivative $\tilde{s}_{;ab}$ and reads
\begin{equation} \label{eq:feature:3}
  {\cal F}^{(3)} = \frac{(\tilde{s}_{;11}-\tilde{s}_{;22}) \tilde{s}_{;1} \tilde{s}_{;2} - \tilde{s}_{;12}(\tilde{s}_{;1}^2 - \tilde{s}_{;2}^2)}{\tilde{s}_{;1}^2 + \tilde{s}_{;2}^2} .
\end{equation}
This expression defines the skeleton of the map $\tilde{s}$, which represents pairs of field lines connecting saddle points to local maxima and minima, initially aligned with the major axis of local curvature \citep{2006MNRAS.366.1201N}.

All these features are determined from the noisy data $d$, and as such are subject to noise. Contribution of small Gaussian noise component to these can be approximated by linear perturbations, covariance of which can be computed analytically. This sets natural scale at which to consider features as ``similar'', or more precisely defines a metric on the feature space. Details of this calculation are presented in Appendix~\ref{sec:app:variance}. For our feature choice, covariance matrix is diagonal, which simplifies things.

\section{Planck 353GHz maps}
\label{sec:dust}

\begin{figure*}
    \centering
    \includegraphics{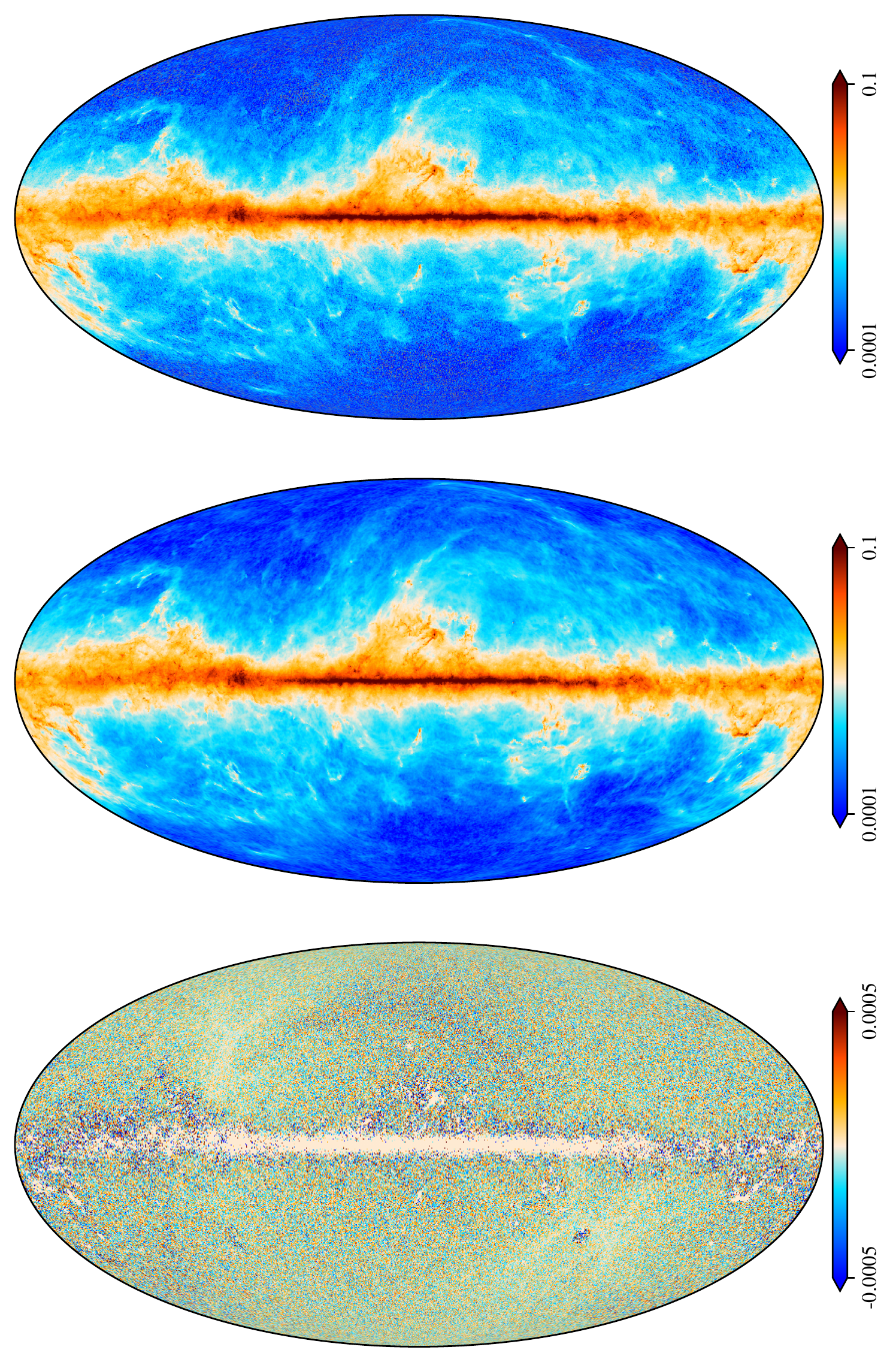}
    \caption{Application of the non-local means filter to a thermal dust emission map at $353$GHz, with resolution $\nside=2048$. Units are $K_{\text{CMB}}$. \emph{Upper:} Intensity channel of a thermal dust emission map is the input map. \emph{Middle:} The output map obtained by non-local means filtering, using a $20'$ FWHM smoothing for the feature space construction and a filtering parameter $\alpha=16$. \emph{Lower:} The difference between the input and output maps showing what was removed by the filter, which we will call a residual map.}
    \label{fig:dust}
\end{figure*}

The thermal emission caused by interstellar dust grains is a significant diffuse astrophysical foreground that introduces contamination to the CMB, which is significantly polarized \citep{2020A&A...641A..11P}. The emission from interstellar dust grains is typically modelled by a modified black body spectrum, with grain temperature around $20-30$K, and power law exponent of around $1.5$ \citep{2014A&A...571A..11P, 2015A&A...576A.107P}. It reaches its peak intensity at very high frequencies (above $1000$GHz), yet it remains a substantial component of the signal within the $30-300$GHz range, which is the frequency range where measurements of CMB anisotropies are performed. 

The Planck mission conducted extensive all-sky measurements in eleven frequency bands covering a range from $30$ to $857$GHz, capturing data on intensity and polarization \citep{2020A&A...641A...1P}. However, polarization sensitivity is limited to frequencies up to $353$GHz. As the frequency increases, the intensity of the CMB and synchrotron radiation diminishes, and thermal dust emission becomes the dominant signal, especially in polarization. At frequencies above $100$GHz, total intensity measurements are primarily influenced by the galactic thermal dust emission and the cosmic infrared background. Similarly, total polarization measurements are predominantly governed by the effects of thermal dust emission.

\begin{figure*}
    \centering
    \includegraphics{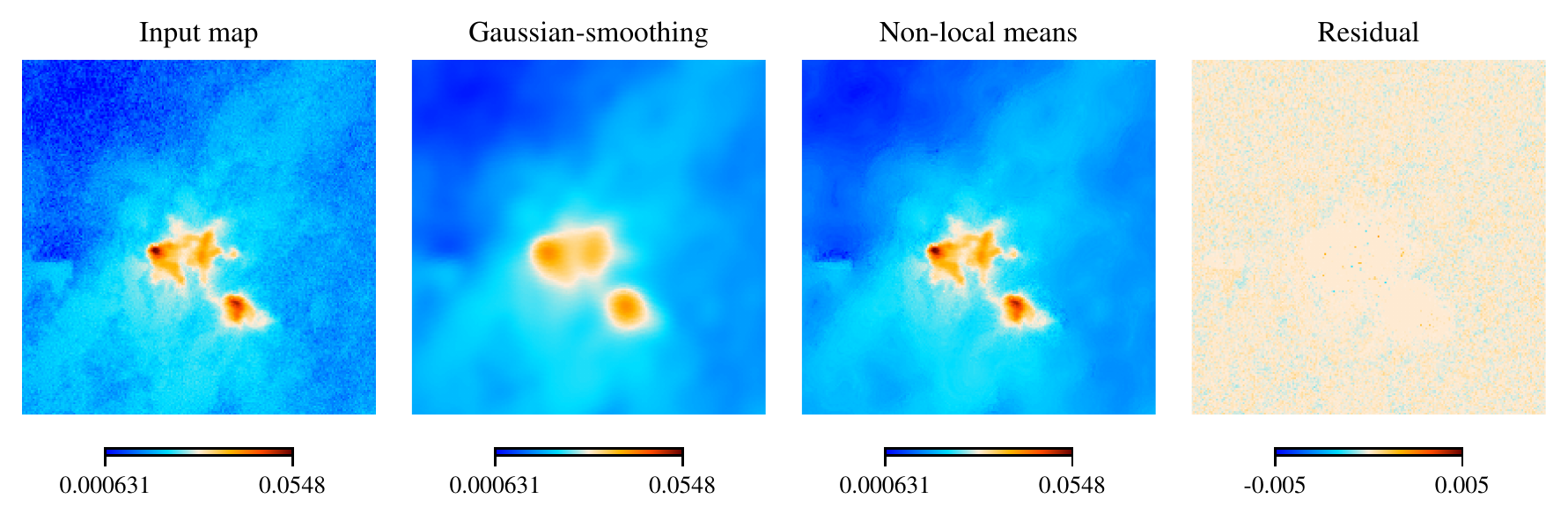}
    \caption{Gnomonic projection of a neighbourhood of the two complex-shaped hot spots for visual comparison. The Gaussian-smoothed map was obtained with a smoothing parameter $\mathrm{FWHM}=20'$. Units are $K_{\text{CMB}}$.}
    \label{fig:dust:gnomonic}
\end{figure*}

Among various channels of the Planck mission, the $353$GHz one stands out as the most sensitive for capturing and studying polarized thermal dust emission. This particular frequency band enables detailed investigations into the properties, distribution, and behavior of dust particles within our galactic environment \citep{2020A&A...641A..12P}.

The Planck mission has released sky maps in 2013, 2015, and 2018 \citep{2014A&A...571A...1P, 2016A&A...594A...1P, 2020A&A...641A...2P, 2020A&A...641A...3P}. The data products include full-mission maps, half-mission and even-odd stripe splits of the data with uncorrelated noise and largely uncorrelated systematics, as well as particular detector sets in the earlier releases. Maps are provided in HEALPix format \citep{2005ApJ...622..759G}, most at $\nside=2048$ and effective Gaussian beam of $5'$ FWHM. Details are explained in \href{https://wiki.cosmos.esa.int/planck-legacy-archive/index.php/Main_Page}{Planck Explanatory Supplement}, in particular \href{https://wiki.cosmos.esa.int/planck-legacy-archive/index.php/Frequency_maps}{Frequency maps section}. In this paper, we use minimally processed $353$GHz map (with CMB component and point sources subtracted as explained in Appendix~\ref{sec:app:dust}) as an example of strongly non-homogeneous and non-Gaussian foreground to test the efficiency of the non-local means noise reduction algorithm.

Fig.~\ref{fig:dust} displays the results of our modified non-local means algorithm applied to a minimally processed full mission Planck $353$GHz intensity map (a good proxy for thermal dust emission) contaminated with noise. The top image illustrates the original map, which exhibits noticeable noise particularly in the high latitude regions. In the middle image, we present the output of our algorithm, demonstrating noise reduction and improved image quality. Finally, the bottom image shows the difference between input and output maps representing what was removed by the filter. Visually, this difference map (which we will call the residual) appears to be mostly noise in regions with low signal-to-noise ratios, and gradually fading to zero towards the galactic plane where the signal-to-noise ratio is significantly higher. This is due to the fact that similarity distance in equation (\ref{eq:weight}) between different bright pixels is large, and weight function effectively concentrates on a single pixel.

Fig.~\ref{fig:dust:gnomonic} presents gnomonic projection of these maps at a specific location to facilitate visual comparison. The second image shows an additional map smoothed with a 20' FWHM Gaussian kernel for reference. It is observed that our proposed non-local means algorithm effectively eliminates noise while preserving the morphological characteristics of the original map, with no apparent loss in resolution. In contrast, convolution with a Gaussian kernel noticeably smoothes the map and alters the shape of the hot spots.

\begin{figure}
    \centering
    \includegraphics[width=\columnwidth]{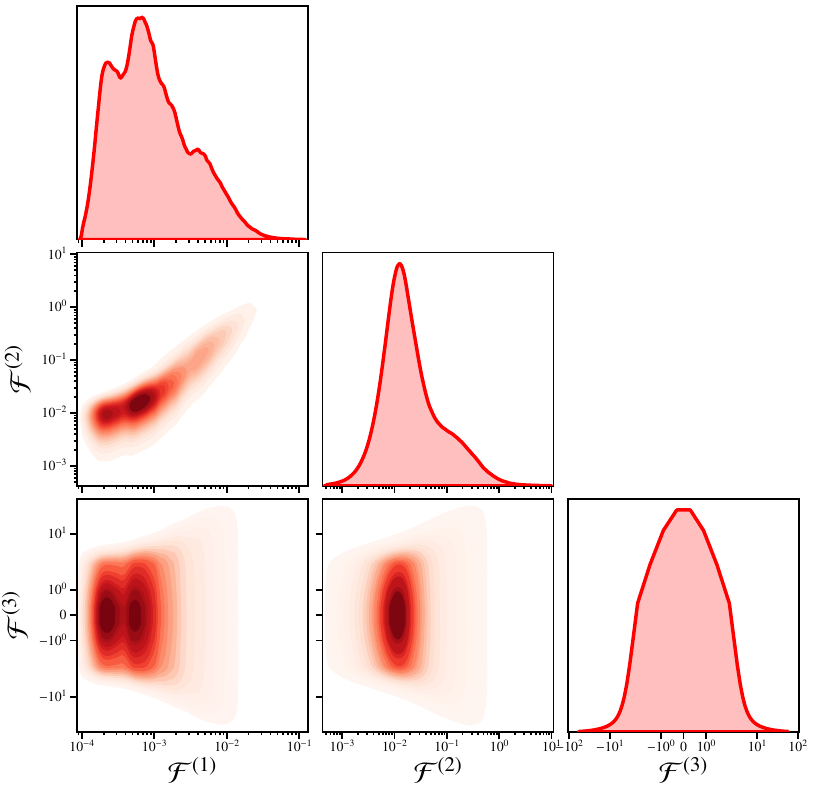}
    \caption{Pairwise bivariate distributions of the feature space components in the lower triangle and marginal distribution of each feature in the feature space on the diagonal.}
    \label{fig:distributions}
\end{figure}

The three features described in Section \ref{sec:feature} were employed to obtain the aforementioned results. One might ask how they characterize the underlying signal itself. For example, Minkowski functionals were used to characterize non-Gaussianity of CMB maps \citep{1998MNRAS.297..355S}. Situation with foregrounds is more complex. Fig.~\ref{fig:distributions} displays the bivariate distributions of these features for the dust map, illustrating a non-trivial correlation between them. Additionally, the marginal distributions of each feature demonstrate distinct characteristics: ${\cal F}^{(1)}$ and ${\cal F}^{(2)}$ are closer to log-normal distribution (actually still skewed even on logarithmic scale), while ${\cal F}^{(3)}$ more resembles a centered normal distribution (with some kurtosis).

\begin{figure}
    \centering
    \includegraphics{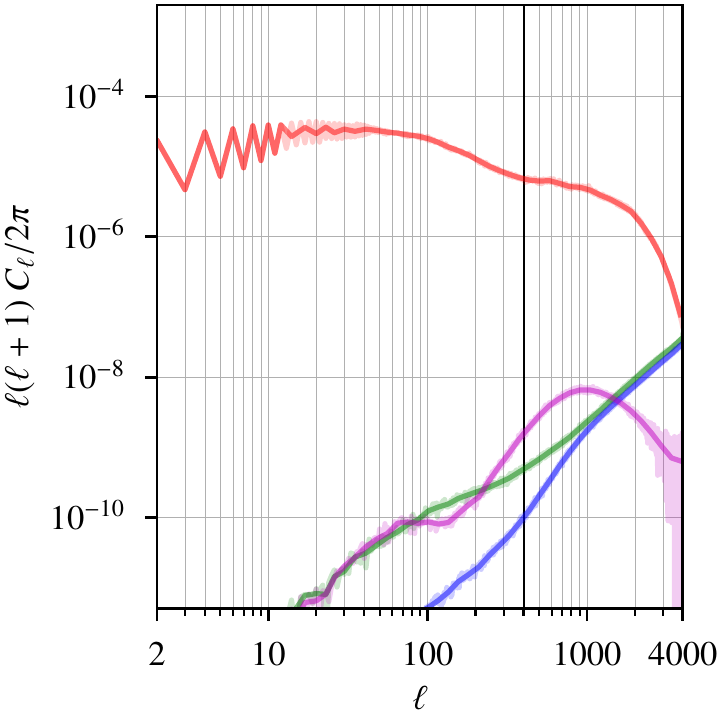}
    \caption{Input signal (red), input noise (green), lost signal (purple) and removed noise (blue) power spectra for even-odd split of the full resolution dust intensity map with $\nside=2048$. Vertical black line corresponds to feature space smoothing scale.}
    \label{fig:dust:spectra}
\medskip
    \centering
    \includegraphics{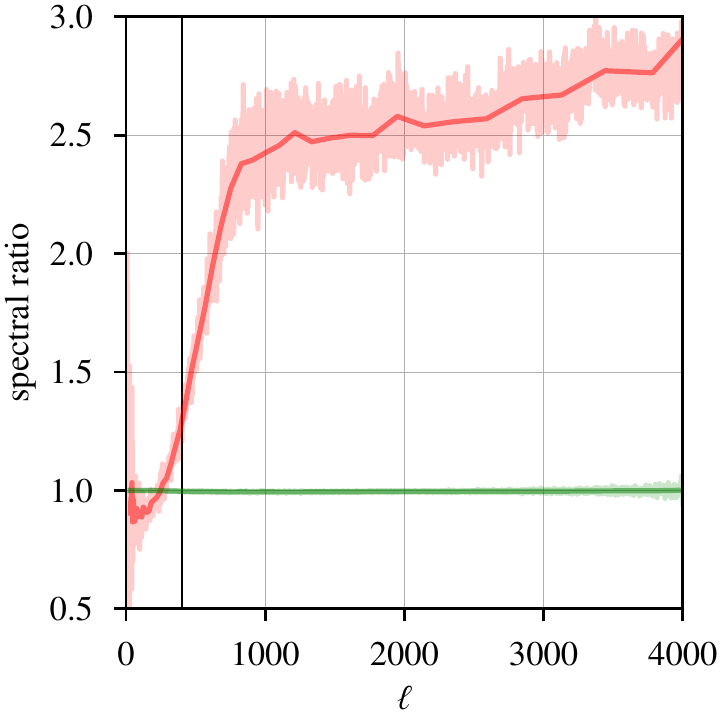}
    \caption{Spectral density signal to noise enhancement (red) and signal attenuation (green) for the full resolution dust intensity map ($\nside=2048$). Vertical black line corresponds to feature space smoothing scale.}
    \label{fig:dust:sn}
\end{figure}

To quantitatively assess the effectiveness of our proposed algorithm, one can independently apply the non-local means filter to odd and even splits of the Planck 353 GHz thermal dust emission maps. The cross-spectrum $C^{\rm OE}_{\rm input}$ of these splits enables us to characterize the true clean signal in input maps
\begin{equation}
    C_{\ell, \, \rm clean} = C^{\rm OE}_{\ell, \, \rm input} ,
\end{equation}
while excess power in autocorrelation of each split contains specific noise contributions (and residual systematics which are lower in the odd-even split than in the half-mission one). We can estimate the power spectrum associated with the noise in input maps by subtracting the cross-spectrum $C^{\rm OE}_{\ell, \, \rm input}$ from the average power spectrum of the splits
\begin{equation}
    C_{\ell, \, \rm noise} = \frac{C^{\rm O}_{\ell, \, \rm input} + C^{\rm E}_{\ell, \, \rm input}}{2} - C^{\rm OE}_{\ell, \, \rm input}.
\end{equation}
The residuals removed by the algorithm may not be perfect and could include contributions from the true signal, which is an undesirable but often unavoidable effect of any filter. To characterize the power spectrum of the lost signal, cross-spectrum of the residuals removed from odd and even maps can be considered
\begin{equation}
    C_{\ell, \, \rm lost} = C^{\rm OE}_{\ell, \, \rm residual}.
\end{equation}
To estimate the noise that has been removed during the process, we subtract the cross-spectrum $C^{\rm OE}_{\ell, \, \rm residual}$ from the average power spectrum of the residuals
\begin{equation}
    C_{\ell, \, \rm removed} = \frac{C^{\rm O}_{\ell, \, \rm residual} + C^{\rm E}_{\ell, \, \rm residual}}{2} - C^{\rm OE}_{\ell, \, \rm residual}.
\end{equation}
The four spectra mentioned are presented in Fig.~\ref{fig:dust:spectra} on a logarithmic scale (evaluated for the full sky coverage). Inspecting the plot, it is evident that the power spectra of the true clean signal are higher than those of the true noise, which means Planck $353$GHz map signal to noise ratio is pretty high as is. Lost signal power is orders of magnitude below the signal, so filtering has minimal impact on the signal. Furthermore, as the multipole moment $\ell$ increases, the spectrum of the removed noise progressively approaches the spectrum of the true noise.

To quantify improvement in the signal to noise ratio due to the filter applied, we can extract signal and noise power spectra from the output maps exactly as we did with input ones, namely
\begin{equation}
    C^{\prime}_{\ell, \, \rm clean} = C^{\rm OE}_{\ell, \, \rm output}
\end{equation}
and
\begin{equation}
    C^{\prime}_{\ell, \, \rm noise} = \frac{C^{\rm O}_{\ell, \, \rm output} + C^{\rm E}_{\ell, \, \rm output}}{2} - C^{\rm OE}_{\ell, \, \rm output} .
\end{equation}
The spectral density signal to noise (SN) ratio for the original data can be expressed as
\begin{equation}
    \text{SN}_\ell = \frac{C_{\ell, \, \rm clean}}{C_{\ell, \, \rm noise}} ,
\end{equation}
while for the filtered maps it is
\begin{equation}
    \text{SN}^{\prime}_\ell = \frac{C^{\prime}_{\ell, \, \rm clean}}{C^{\prime}_{\ell, \, \rm noise}} .
\end{equation}
Fig.~\ref{fig:dust:sn} depicts the enhancement of the signal to noise ratio, quantified by ${\text{SN}^{\prime}_\ell} / {\text{SN}_\ell}$, and the signal attenuation, quantified by the ratio $C_{\ell, \, \text{clean}}^{\prime} / C_{\ell, \, \text{clean}}$. It can be observed that our non-local means algorithm achieves a significant spectral SN enhancement, which increases at higher multipole moment $\ell$. Additionally, the signal attenuation remains negligible across all scales. For comparison, an optimal linear filter would attenuate signal and noise spectral densities equally, resulting in spectral signal to noise ratio of one, with any gains realized only in integrated signal.

\section{Component-separated CMB maps}
\label{sec:cmb}

\begin{figure}
    \centering
    \includegraphics{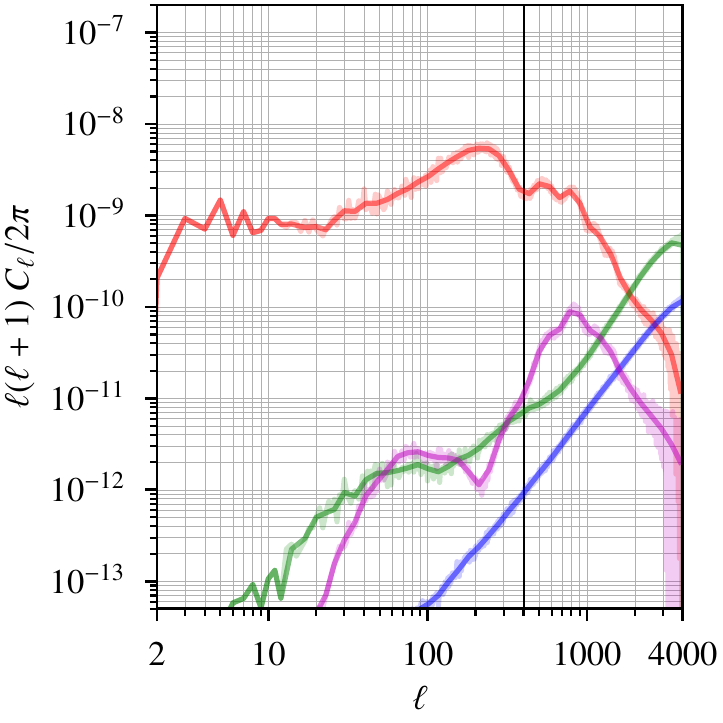}
    \caption{Input signal (red), input noise (green), lost signal (purple) and removed noise (blue) power spectra for even-odd split of the full resolution CMB component separated map with $\nside=2048$. Vertical black line corresponds to feature space smoothing scale.}
    \label{fig:cmb:spectra}
\end{figure}

We also evaluated non-local means filter for noise reduction in Planck component separated maps \citep{2020A&A...641A...4P}. Planck provides component separation by four different methods, known as SMICA, SEVEM, NILC, and Commander. Commander uses spectral energy density models for foregrounds to evaluate best fit for components on per-pixel basis, while the other three methods are based on various linear combination strategies. For our test map, we use 2018 SMICA component separated CMB map, which is supplied at $\nside=2048$ resolution with 5' FWHM Gaussian beam. As other Planck data products, it is available as splits as well.

We tried several smoothing scales for feature space construction and different filter strengths. Fig.~\ref{fig:cmb:spectra} shows the four spectral densities we used to characterize dust map filtering, as described in the previous section, for 20' FWHM Gaussian smoothing scale with filter strength $\alpha=32$. Red and green curves represent input signal and noise spectra, while purple and blue ones show removed signal and noise. Unlike dust map, signal and noise at higher $\ell$ are removed with about the same efficiency, which means there is little gain in spectral signal to noise ratio. In this sense, the non-local means filter performance is not much better than the linear filter (for example matched Wiener filter) could achieve at less computational expense.

The reason for this is simple. CMB temperature anisotropy is an isotropic Gaussian random field, and the only thing distinguishing it from noise is different spectra, which are hard to disentangle from a single map. Unlike dust emission which has prominent features the non-local means filter can use to separate signal from noise components based on morphology, CMB features are not as distinctive. Non-local means filter still reduces noise, of course, but struggles to separate signal from noise based on morphology only. This does not mean it cannot improve component separated CMB maps, but the way to do it would be to remove noise from \textit{foreground} maps used as input for component separation. Reduced noise in foreground templates used to extract CMB signal would directly translate into reduced noise of the linear combination map. In many ILC strategies, it would also help with determining linear weights more accurately, as it would tighten up covariance matrices used to determine them.

\section{Extensions to polarization}
\label{sec:polar}

Polarization measurements present much larger possibilities for feature space construction, while coming with a number of specific challenges related to how the polarization data is used for scientific inference in cosmology. Unlike scalar intensity maps we discussed up to this point, linearly polarized emission is described by a rank two tensor, with components in local orthonormal frame represented by Stokes parameters $I$, $Q$ and $U$ as
\begin{equation}
  {\cal P}_{ab} = \left[\begin{array}{cc} I+Q & U\\ U & I-Q\end{array}\right] .
\end{equation}
Intensity $I = \frac{1}{2}\, {\cal P}_a^a$ is a scalar, as well as total polarization power $P^2 = \frac{1}{2}\, P_{ab} P^{ab} = Q^2 + U^2$ constructed from traceless tensor $P_{ab} \equiv {\cal P}_{ab} - I\, \delta_{ab}$ describing purely polarized component. A number of invariants involving derivatives can be readily constructed, for example $P^{ab}I_{;a}I_{;b}$, $P^{ab}I_{;a}\epsilon_{bc}I^{;c}$, $P^{ab}\epsilon_{ac}I^{;c}\epsilon_{bd}I^{;d}$, as well as higher order ones like $P^{ab}_{~~;ab}$, $P^{ab;c}_{~~~;a}\epsilon_{bc}$ involving polarization only, or coupling to intensity, e.g.\ $P^{ab}I_{;ab}$. Among the multitude of choices that could be used for feature space construction, one should be aware that non-intended correlations could be introduced by the selection. While equation (\ref{eq:nlmean}) looks data-linear, it is not as weights (\ref{eq:weight}) depend on data too. If one is not careful, correlations or mode conversion between intensity and two parity modes of polarization could occur. This would not be a big problem if all signals had comparable power, but for cosmological maps, intensity of CMB is an order of magnitude larger than even parity $E$-mode, which in turn has more power than odd parity $B$-mode (exactly how much depends on the inflation model), detecting which is the point of many future CMB studies.

Additional complication arises from the fact that Stokes parameters $Q$ and $U$ are really projections onto a local coordinate frame, and averaging them between different pixels is highly non-trivial. The data sets provided by the Planck collaboration utilize the HEALPix pixelization scheme to store polarized data, with local orthonormal frame aligned with $\hat{\theta}$ and $\hat{\phi}$ directions. These could be vastly different even for neighbouring pixels, for example around the poles. While one could rotate the frames of nearby pixels to align in a tangent space, this is by no means easy to implement \citep{2020A&A...641A...7P}, and would not work for global averages anyway. A better approach is to operate on parity-definite scalars constructed from polarization data \citep{1997PhRvD..55.1830Z}.

A useful representation of polarization is in terms of (complex) spin-weighted spherical harmonics. Consider the fields $Q(\hat{\vec{n}})$ and $U(\hat{\vec{n}})$ representing the Stokes parameters $Q$ and $U$ in the direction $\hat{\vec{n}}$. From these fields, one can construct two spin-2 fields, namely $Q + iU$ and $Q - iU$. In terms of the spin-2 weighted spherical harmonics $_{\pm 2}Y_{\ell m}$, these fields can be expressed as
\begin{equation}
    (Q \pm iU)(\hat{\vec{n}}) = \sum_{\ell m} a_{\pm 2,\ell m} \; _{\pm 2}Y_{\ell m}(\hat{\vec{n}}) .
\end{equation}
To extend our denoising algorithm for polarization maps, one option is to directly apply it to spin-0 quantities
\begin{equation} \label{eq:eth}
    \eth^{2}_{\mp} (Q \pm iU)(\hat{\vec{n}}) = \sum_{\ell m} \sqrt{\frac{(l+2)!}{(l-2)!}} \; a_{\pm 2,\ell m} \; Y_{\ell m}(\hat{\vec{n}}) ,
\end{equation}
which correspond to two second derivative differential invariants of polarization tensor mentioned above. These definitions are local, and can be evaluated on a masked sky. However, their anisotropy spectra have contribution of two derivatives, emphasizing high-frequency noise.

More convenient variables are known as $E$- and $B$-modes, which are essentially inverse Laplacian of equation (\ref{eq:eth}). They can be expressed as spherical harmonic expansions in terms of the coefficients $a_{\pm 2,\ell m}$ as
\begin{equation}
    \begin{aligned}
        E(\hat{\vec{n}}) = \sum_{\ell m} a_{E,\ell m} Y_{\ell m}(\hat{\vec{n}}) , \\
        B(\hat{\vec{n}}) = \sum_{\ell m} a_{B,\ell m} Y_{\ell m}(\hat{\vec{n}}) ,
    \end{aligned}
\end{equation}
where
\begin{equation}
    \begin{aligned}
        a_{E,\ell m} =  -\frac{a_{+2,\ell m} + a_{-2,\ell m}}{2} , \\
        a_{B,\ell m} = -\frac{a_{+2,\ell m} - a_{-2,\ell m}}{2i} .
    \end{aligned}
\end{equation}
These maps are rotationally invariant. Specifically, $E$ behaves as a scalar field, while $B$ behaves as a pseudo-scalar field. Similar to our study in Section \ref{sec:feature}, the feature space can be constructed using these rotationally invariant maps, from the associated Gaussian smoothed maps $\tilde{E}$ and $\tilde{B}$, as well as other invariant quantities that involve covariant derivatives, such as $|\nabla \tilde{E}|$ or $|\nabla \tilde{B}|$. The necessity and impact of inclusion of higher derivatives in the polarization feature space construction bears further investigation.

\section{Conclusions}
\label{sec:conclusion}

In this paper, we discuss a new non-linear noise reduction algorithm for scalar data on a sphere, and its implementation for HEALPix pixelization of the maps used in cosmic microwave background anisotropy studies. It is based on ideas of non-local means algorithm in digital signal processing by \cite{Buades:2005}, but is specifically adopted to the symmetries of the CMB and astrophysical foreground maps. The noise is removed by averaging ``similar'' pixels, with similarity determined by a tower of differential invariants forming a feature space, outfitted with a distance measure calculated from the noise covariance of the feature estimators.

The algorithm is substantially more effective than anything else we are aware of for non-Gaussian emission maps, realizing a factor of two gain in spectral signal to noise ratio without any apparent signal loss for Planck $353$GHz dust maps. Application to component separated CMB maps is less spectacular, with efficiency roughly comparable to linear filters. Although we mostly focused on emission intensity maps, the same techniques can be applied to polarization data, with potentially larger feature space constructed from polarization tensor. To avoid unintended correlations and mode conversions, it seems prudent to apply the filter separately on parity-definite scalar maps, namely $E$- and $B$-modes. These are easily constructed from full-sky maps, but are far less trivial to estimate on a masked sky.

Impact of the noise reduction is more apparent for the astrophysical foreground maps, which have strong features the algorithm can use to separate signal from noise based on morphology only, without resorting to frequency dependence of emission. This is advantageous especially in the context of ``spectral confusion'', i.e.\ when foregrounds are hard to disentangle on their spectral energy density profiles alone, which is an ever-present worry for component separation techniques. Even for clearly spectrally distinguished foregrounds, reducing the noise in foreground templates would correspondingly decrease it in component-separated CMB maps. Given that a factor of two is feasible at least for some maps with moderate computational expenses (which would increase hardware cost or integration time by a factor of four if brute-force data accumulation strategy was to be used to reduce statistical noise), it seems like a promising technique to explore.

Of potential downsides, the non-linear nature of the filter might complicate statistical analysis. However, the foreground properties and instrumental noise models are already complicated as they are, and frequentist approach with full forward-feeding signal and noise simulations is already used in Planck. This trend might continue in future experiments, opening the window of opportunity for non-linear signal processing. It is already widely used in image processing and computer vision.

\begin{acknowledgements}
This work was supported in part by NSERC Discovery Grant ``Testing fundamental physics with B-modes of Cosmic Microwave Background anisotropy''.
\end{acknowledgements}

\bibliographystyle{aa}
\bibliography{references,planck,cmb,compsep}

\begin{thebibliography}{36}
\expandafter\ifx\csname natexlab\endcsname\relax\def\natexlab#1{#1}\fi

\bibitem[{{Ade} {et~al.}(2021){Ade}, {Ahmed}, {Amiri}, {Barkats}, {Thakur},
  {Bischoff}, {Beck}, {Bock}, {Boenish}, {Bullock}, {Buza}, {Cheshire},
  {Connors}, {Cornelison}, {Crumrine}, {Cukierman}, {Denison}, {Dierickx},
  {Duband}, {Eiben}, {Fatigoni}, {Filippini}, {Fliescher}, {Goeckner-Wald},
  {Goldfinger}, {Grayson}, {Grimes}, {Hall}, {Halal}, {Halpern}, {Hand},
  {Harrison}, {Henderson}, {Hildebrandt}, {Hilton}, {Hubmayr}, {Hui}, {Irwin},
  {Kang}, {Karkare}, {Karpel}, {Kefeli}, {Kernasovskiy}, {Kovac}, {Kuo}, {Lau},
  {Leitch}, {Lennox}, {Megerian}, {Minutolo}, {Moncelsi}, {Nakato}, {Namikawa},
  {Nguyen}, {O'Brient}, {Ogburn}, {Palladino}, {Prouve}, {Pryke}, {Racine},
  {Reintsema}, {Richter}, {Schillaci}, {Schwarz}, {Schmitt}, {Sheehy},
  {Soliman}, {Germaine}, {Steinbach}, {Sudiwala}, {Teply}, {Thompson}, {Tolan},
  {Tucker}, {Turner}, {Umilt{\`a}}, {Verg{\`e}s}, {Vieregg}, {Wandui}, {Weber},
  {Wiebe}, {Willmert}, {Wong}, {Wu}, {Yang}, {Yoon}, {Young}, {Yu}, {Zeng},
  {Zhang}, {Zhang}, \& {Bicep/Keck Collaboration}}]{2021PhRvL.127o1301A}
{Ade}, P.~A.~R., {Ahmed}, Z., {Amiri}, M., {et~al.} 2021, \prl, 127, 151301

\bibitem[{{Bond} {et~al.}(1994){Bond}, {Crittenden}, {Davis}, {Efstathiou}, \&
  {Steinhardt}}]{1994PhRvL..72...13B}
{Bond}, J.~R., {Crittenden}, R., {Davis}, R.~L., {Efstathiou}, G., \&
  {Steinhardt}, P.~J. 1994, \prl, 72, 13

\bibitem[{{Bond} \& {Efstathiou}(1987)}]{1987MNRAS.226..655B}
{Bond}, J.~R. \& {Efstathiou}, G. 1987, \mnras, 226, 655

\bibitem[{{Brandt}(1977)}]{Brandt:1977}
{Brandt}, A. 1977, Mathematics of Computation, 31, 333

\bibitem[{Buades {et~al.}(2005)Buades, Coll, \& Morel}]{Buades:2005}
Buades, A., Coll, B., \& Morel, J.-M. 2005, in IN CVPR, 60--65

\bibitem[{{Cardoso} {et~al.}(2008){Cardoso}, {Martin}, {Delabrouille},
  {Betoule}, \& {Patanchon}}]{2008arXiv0803.1814C}
{Cardoso}, J.-F., {Martin}, M., {Delabrouille}, J., {Betoule}, M., \&
  {Patanchon}, G. 2008, arXiv e-prints, arXiv:0803.1814

\bibitem[{{Crittenden} {et~al.}(1993){Crittenden}, {Bond}, {Davis},
  {Efstathiou}, \& {Steinhardt}}]{1993PhRvL..71..324C}
{Crittenden}, R., {Bond}, J.~R., {Davis}, R.~L., {Efstathiou}, G., \&
  {Steinhardt}, P.~J. 1993, \prl, 71, 324

\bibitem[{{Delabrouille} {et~al.}(2003){Delabrouille}, {Cardoso}, \&
  {Patanchon}}]{2003MNRAS.346.1089D}
{Delabrouille}, J., {Cardoso}, J.~F., \& {Patanchon}, G. 2003, \mnras, 346,
  1089

\bibitem[{{Eriksen} {et~al.}(2008){Eriksen}, {Jewell}, {Dickinson}, {Banday},
  {G{\'o}rski}, \& {Lawrence}}]{2008ApJ...676...10E}
{Eriksen}, H.~K., {Jewell}, J.~B., {Dickinson}, C., {et~al.} 2008, \apj, 676,
  10

\bibitem[{{G{\'o}rski} {et~al.}(2005){G{\'o}rski}, {Hivon}, {Banday},
  {Wandelt}, {Hansen}, {Reinecke}, \& {Bartelmann}}]{2005ApJ...622..759G}
{G{\'o}rski}, K.~M., {Hivon}, E., {Banday}, A.~J., {et~al.} 2005, \apj, 622,
  759

\bibitem[{{Hinshaw} {et~al.}(2013){Hinshaw}, {Larson}, {Komatsu}, {Spergel},
  {Bennett}, {Dunkley}, {Nolta}, {Halpern}, {Hill}, {Odegard}, {Page}, {Smith},
  {Weiland}, {Gold}, {Jarosik}, {Kogut}, {Limon}, {Meyer}, {Tucker}, {Wollack},
  \& {Wright}}]{2013ApJS..208...19H}
{Hinshaw}, G., {Larson}, D., {Komatsu}, E., {et~al.} 2013, \apjs, 208, 19

\bibitem[{{Leach} {et~al.}(2008){Leach}, {Cardoso}, {Baccigalupi}, {Barreiro},
  {Betoule}, {Bobin}, {Bonaldi}, {Delabrouille}, {de Zotti}, {Dickinson},
  {Eriksen}, {Gonz{\'a}lez-Nuevo}, {Hansen}, {Herranz}, {Le Jeune},
  {L{\'o}pez-Caniego}, {Mart{\'\i}nez-Gonz{\'a}lez}, {Massardi}, {Melin},
  {Miville-Desch{\^e}nes}, {Patanchon}, {Prunet}, {Ricciardi}, {Salerno},
  {Sanz}, {Starck}, {Stivoli}, {Stolyarov}, {Stompor}, \&
  {Vielva}}]{2008A&A...491..597L}
{Leach}, S.~M., {Cardoso}, J.~F., {Baccigalupi}, C., {et~al.} 2008, \aap, 491,
  597

\bibitem[{{Mart{\'\i}nez-Gonz{\'a}lez}
  {et~al.}(2003){Mart{\'\i}nez-Gonz{\'a}lez}, {Diego}, {Vielva}, \&
  {Silk}}]{2003MNRAS.345.1101M}
{Mart{\'\i}nez-Gonz{\'a}lez}, E., {Diego}, J.~M., {Vielva}, P., \& {Silk}, J.
  2003, \mnras, 345, 1101

\bibitem[{{Novikov} {et~al.}(2006){Novikov}, {Colombi}, \&
  {Dor{\'e}}}]{2006MNRAS.366.1201N}
{Novikov}, D., {Colombi}, S., \& {Dor{\'e}}, O. 2006, \mnras, 366, 1201

\bibitem[{{Planck Collaboration} {et~al.}(2014{\natexlab{a}}){Planck
  Collaboration}, {Abergel}, {Ade}, {Aghanim}, {Alves}, {Aniano},
  {Armitage-Caplan}, {Arnaud}, {Ashdown}, {Atrio-Barandela}, {Aumont},
  {Baccigalupi}, {Banday}, {Barreiro}, {Bartlett}, {Battaner}, {Benabed},
  {Beno{\^\i}t}, {Benoit-L{\'e}vy}, {Bernard}, {Bersanelli}, {Bielewicz},
  {Bobin}, {Bock}, {Bonaldi}, {Bond}, {Borrill}, {Bouchet}, {Boulanger},
  {Bridges}, {Bucher}, {Burigana}, {Butler}, {Cardoso}, {Catalano},
  {Chamballu}, {Chary}, {Chiang}, {Chiang}, {Christensen}, {Church}, {Clemens},
  {Clements}, {Colombi}, {Colombo}, {Combet}, {Couchot}, {Coulais}, {Crill},
  {Curto}, {Cuttaia}, {Danese}, {Davies}, {Davis}, {de Bernardis}, {de Rosa},
  {de Zotti}, {Delabrouille}, {Delouis}, {D{\'e}sert}, {Dickinson}, {Diego},
  {Dole}, {Donzelli}, {Dor{\'e}}, {Douspis}, {Draine}, {Dupac}, {Efstathiou},
  {En{\ss}lin}, {Eriksen}, {Falgarone}, {Finelli}, {Forni}, {Frailis},
  {Fraisse}, {Franceschi}, {Galeotta}, {Ganga}, {Ghosh}, {Giard}, {Giardino},
  {Giraud-H{\'e}raud}, {Gonz{\'a}lez-Nuevo}, {G{\'o}rski}, {Gratton},
  {Gregorio}, {Grenier}, {Gruppuso}, {Guillet}, {Hansen}, {Hanson}, {Harrison},
  {Helou}, {Henrot-Versill{\'e}}, {Hern{\'a}ndez-Monteagudo}, {Herranz},
  {Hildebrandt}, {Hivon}, {Hobson}, {Holmes}, {Hornstrup}, {Hovest},
  {Huffenberger}, {Jaffe}, {Jaffe}, {Jewell}, {Joncas}, {Jones}, {Juvela},
  {Keih{\"a}nen}, {Keskitalo}, {Kisner}, {Knoche}, {Knox}, {Kunz},
  {Kurki-Suonio}, {Lagache}, {L{\"a}hteenm{\"a}ki}, {Lamarre}, {Lasenby},
  {Laureijs}, {Lawrence}, {Leonardi}, {Le{\'o}n-Tavares}, {Lesgourgues},
  {Levrier}, {Liguori}, {Lilje}, {Linden-V{\o}rnle}, {L{\'o}pez-Caniego},
  {Lubin}, {Mac{\'\i}as-P{\'e}rez}, {Maffei}, {Maino}, {Mandolesi}, {Maris},
  {Marshall}, {Martin}, {Mart{\'\i}nez-Gonz{\'a}lez}, {Masi}, {Massardi},
  {Matarrese}, {Matthai}, {Mazzotta}, {McGehee}, {Melchiorri}, {Mendes},
  {Mennella}, {Migliaccio}, {Mitra}, {Miville-Desch{\^e}nes}, {Moneti},
  {Montier}, {Morgante}, {Mortlock}, {Munshi}, {Murphy}, {Naselsky}, {Nati},
  {Natoli}, {Netterfield}, {N{\o}rgaard-Nielsen}, {Noviello}, {Novikov},
  {Novikov}, {Osborne}, {Oxborrow}, {Paci}, {Pagano}, {Pajot}, {Paladini},
  {Paoletti}, {Pasian}, {Patanchon}, {Perdereau}, {Perotto}, {Perrotta},
  {Piacentini}, {Piat}, {Pierpaoli}, {Pietrobon}, {Plaszczynski},
  {Pointecouteau}, {Polenta}, {Ponthieu}, {Popa}, {Poutanen}, {Pratt},
  {Pr{\'e}zeau}, {Prunet}, {Puget}, {Rachen}, {Reach}, {Rebolo}, {Reinecke},
  {Remazeilles}, {Renault}, {Ricciardi}, {Riller}, {Ristorcelli}, {Rocha},
  {Rosset}, {Roudier}, {Rowan-Robinson}, {Rubi{\~n}o-Mart{\'\i}n}, {Rusholme},
  {Sandri}, {Santos}, {Savini}, {Scott}, {Seiffert}, {Shellard}, {Spencer},
  {Starck}, {Stolyarov}, {Stompor}, {Sudiwala}, {Sunyaev}, {Sureau}, {Sutton},
  {Suur-Uski}, {Sygnet}, {Tauber}, {Tavagnacco}, {Terenzi}, {Toffolatti},
  {Tomasi}, {Tristram}, {Tucci}, {Tuovinen}, {T{\"u}rler}, {Umana},
  {Valenziano}, {Valiviita}, {Van Tent}, {Verstraete}, {Vielva}, {Villa},
  {Vittorio}, {Wade}, {Wandelt}, {Welikala}, {Ysard}, {Yvon}, {Zacchei}, \&
  {Zonca}}]{2014A&A...571A..11P}
{Planck Collaboration}, {Abergel}, A., {Ade}, P.~A.~R., {et~al.}
  2014{\natexlab{a}}, \aap, 571, A11

\bibitem[{{Planck Collaboration} {et~al.}(2016){Planck Collaboration}, {Adam},
  {Ade}, {Aghanim}, {Akrami}, {Alves}, {Arg{\"u}eso}, {Arnaud}, {Arroja},
  {Ashdown}, {Aumont}, {Baccigalupi}, {Ballardini}, {Banday}, {Barreiro},
  {Bartlett}, {Bartolo}, {Basak}, {Battaglia}, {Battaner}, {Battye}, {Benabed},
  {Beno{\^\i}t}, {Benoit-L{\'e}vy}, {Bernard}, {Bersanelli}, {Bertincourt},
  {Bielewicz}, {Bikmaev}, {Bock}, {B{\"o}hringer}, {Bonaldi}, {Bonavera},
  {Bond}, {Borrill}, {Bouchet}, {Boulanger}, {Bucher}, {Burenin}, {Burigana},
  {Butler}, {Calabrese}, {Cardoso}, {Carvalho}, {Casaponsa}, {Castex},
  {Catalano}, {Challinor}, {Chamballu}, {Chary}, {Chiang}, {Chluba}, {Chon},
  {Christensen}, {Church}, {Clemens}, {Clements}, {Colombi}, {Colombo},
  {Combet}, {Comis}, {Contreras}, {Couchot}, {Coulais}, {Crill}, {Cruz},
  {Curto}, {Cuttaia}, {Danese}, {Davies}, {Davis}, {de Bernardis}, {de Rosa},
  {de Zotti}, {Delabrouille}, {Delouis}, {D{\'e}sert}, {Di Valentino},
  {Dickinson}, {Diego}, {Dolag}, {Dole}, {Donzelli}, {Dor{\'e}}, {Douspis},
  {Ducout}, {Dunkley}, {Dupac}, {Efstathiou}, {Eisenhardt}, {Elsner},
  {En{\ss}lin}, {Eriksen}, {Falgarone}, {Fantaye}, {Farhang}, {Feeney},
  {Fergusson}, {Fernandez-Cobos}, {Feroz}, {Finelli}, {Florido}, {Forni},
  {Frailis}, {Fraisse}, {Franceschet}, {Franceschi}, {Frejsel}, {Frolov},
  {Galeotta}, {Galli}, {Ganga}, {Gauthier}, {G{\'e}nova-Santos}, {Gerbino},
  {Ghosh}, {Giard}, {Giraud-H{\'e}raud}, {Giusarma}, {Gjerl{\o}w},
  {Gonz{\'a}lez-Nuevo}, {G{\'o}rski}, {Grainge}, {Gratton}, {Gregorio},
  {Gruppuso}, {Gudmundsson}, {Hamann}, {Handley}, {Hansen}, {Hanson},
  {Harrison}, {Heavens}, {Helou}, {Henrot-Versill{\'e}},
  {Hern{\'a}ndez-Monteagudo}, {Herranz}, {Hildebrandt}, {Hivon}, {Hobson},
  {Holmes}, {Hornstrup}, {Hovest}, {Huang}, {Huffenberger}, {Hurier},
  {Ili{\'c}}, {Jaffe}, {Jaffe}, {Jin}, {Jones}, {Juvela}, {Karakci},
  {Keih{\"a}nen}, {Keskitalo}, {Khamitov}, {Kiiveri}, {Kim}, {Kisner},
  {Kneissl}, {Knoche}, {Knox}, {Krachmalnicoff}, {Kunz}, {Kurki-Suonio},
  {Lacasa}, {Lagache}, {L{\"a}hteenm{\"a}ki}, {Lamarre}, {Langer}, {Lasenby},
  {Lattanzi}, {Lawrence}, {Le Jeune}, {Leahy}, {Lellouch}, {Leonardi},
  {Le{\'o}n-Tavares}, {Lesgourgues}, {Levrier}, {Lewis}, {Liguori}, {Lilje},
  {Lilley}, {Linden-V{\o}rnle}, {Lindholm}, {Liu}, {L{\'o}pez-Caniego},
  {Lubin}, {Ma}, {Mac{\'\i}as-P{\'e}rez}, {Maggio}, {Maino}, {Mak},
  {Mandolesi}, {Mangilli}, {Marchini}, {Marcos-Caballero}, {Marinucci},
  {Maris}, {Marshall}, {Martin}, {Martinelli}, {Mart{\'\i}nez-Gonz{\'a}lez},
  {Masi}, {Matarrese}, {Mazzotta}, {McEwen}, {McGehee}, {Mei}, {Meinhold},
  {Melchiorri}, {Melin}, {Mendes}, {Mennella}, {Migliaccio}, {Mikkelsen},
  {Millea}, {Mitra}, {Miville-Desch{\^e}nes}, {Molinari}, {Moneti}, {Montier},
  {Moreno}, {Morgante}, {Mortlock}, {Moss}, {Mottet}, {M{\"u}nchmeyer},
  {Munshi}, {Murphy}, {Narimani}, {Naselsky}, {Nastasi}, {Nati}, {Natoli},
  {Negrello}, {Netterfield}, {N{\o}rgaard-Nielsen}, {Noviello}, {Novikov},
  {Novikov}, {Olamaie}, {Oppermann}, {Orlando}, {Oxborrow}, {Paci}, {Pagano},
  {Pajot}, {Paladini}, {Pandolfi}, {Paoletti}, {Partridge}, {Pasian},
  {Patanchon}, {Pearson}, {Peel}, {Peiris}, {Pelkonen}, {Perdereau}, {Perotto},
  {Perrott}, {Perrotta}, {Pettorino}, {Piacentini}, {Piat}, {Pierpaoli},
  {Pietrobon}, {Plaszczynski}, {Pogosyan}, {Pointecouteau}, {Polenta}, {Popa},
  {Pratt}, {Pr{\'e}zeau}, {Prunet}, {Puget}, {Rachen}, {Racine}, {Reach},
  {Rebolo}, {Reinecke}, {Remazeilles}, {Renault}, {Renzi}, {Ristorcelli},
  {Rocha}, {Roman}, {Romelli}, {Rosset}, {Rossetti}, {Rotti}, {Roudier},
  {Rouill{\'e} d'Orfeuil}, {Rowan-Robinson}, {Rubi{\~n}o-Mart{\'\i}n},
  {Ruiz-Granados}, {Rumsey}, {Rusholme}, {Said}, {Salvatelli}, {Salvati},
  {Sandri}, {Sanghera}, {Santos}, {Saunders}, {Sauv{\'e}}, {Savelainen},
  {Savini}, {Schaefer}, {Schammel}, {Scott}, {Seiffert}, {Serra}, {Shellard},
  {Shimwell}, {Shiraishi}, {Smith}, {Souradeep}, {Spencer}, {Spinelli},
  {Stanford}, {Stern}, {Stolyarov}, {Stompor}, {Strong}, {Sudiwala}, {Sunyaev},
  {Sutter}, {Sutton}, {Suur-Uski}, {Sygnet}, {Tauber}, {Tavagnacco}, {Terenzi},
  {Texier}, {Toffolatti}, {Tomasi}, {Tornikoski}, {Tramonte}, {Tristram},
  {Troja}, {Trombetti}, {Tucci}, {Tuovinen}, {T{\"u}rler}, {Umana},
  {Valenziano}, {Valiviita}, {Van Tent}, {Vassallo}, {Vibert}, {Vidal}, {Viel},
  {Vielva}, {Villa}, {Wade}, {Walter}, {Wandelt}, {Watson}, {Wehus},
  {Welikala}, {Weller}, {White}, {White}, {Wilkinson}, {Yvon}, {Zacchei},
  {Zibin}, \& {Zonca}}]{2016A&A...594A...1P}
{Planck Collaboration}, {Adam}, R., {Ade}, P.~A.~R., {et~al.} 2016, \aap, 594,
  A1

\bibitem[{{Planck Collaboration} {et~al.}(2014{\natexlab{b}}){Planck
  Collaboration}, {Ade}, {Aghanim}, {Alves}, {Armitage-Caplan}, {Arnaud},
  {Ashdown}, {Atrio-Barandela}, {Aumont}, {Aussel}, {Baccigalupi}, {Banday},
  {Barreiro}, {Barrena}, {Bartelmann}, {Bartlett}, {Bartolo}, {Basak},
  {Battaner}, {Battye}, {Benabed}, {Beno{\^\i}t}, {Benoit-L{\'e}vy}, {Bernard},
  {Bersanelli}, {Bertincourt}, {Bethermin}, {Bielewicz}, {Bikmaev},
  {Blanchard}, {Bobin}, {Bock}, {B{\"o}hringer}, {Bonaldi}, {Bonavera}, {Bond},
  {Borrill}, {Bouchet}, {Boulanger}, {Bourdin}, {Bowyer}, {Bridges}, {Brown},
  {Bucher}, {Burenin}, {Burigana}, {Butler}, {Calabrese}, {Cappellini},
  {Cardoso}, {Carr}, {Carvalho}, {Casale}, {Castex}, {Catalano}, {Challinor},
  {Chamballu}, {Chary}, {Chen}, {Chiang}, {Chiang}, {Chon}, {Christensen},
  {Churazov}, {Church}, {Clemens}, {Clements}, {Colombi}, {Colombo}, {Combet},
  {Comis}, {Couchot}, {Coulais}, {Crill}, {Cruz}, {Curto}, {Cuttaia}, {Da
  Silva}, {Dahle}, {Danese}, {Davies}, {Davis}, {de Bernardis}, {de Rosa}, {de
  Zotti}, {D{\'e}chelette}, {Delabrouille}, {Delouis}, {D{\'e}mocl{\`e}s},
  {D{\'e}sert}, {Dick}, {Dickinson}, {Diego}, {Dolag}, {Dole}, {Donzelli},
  {Dor{\'e}}, {Douspis}, {Ducout}, {Dunkley}, {Dupac}, {Efstathiou}, {Elsner},
  {En{\ss}lin}, {Eriksen}, {Fabre}, {Falgarone}, {Falvella}, {Fantaye},
  {Fergusson}, {Filliard}, {Finelli}, {Flores-Cacho}, {Foley}, {Forni},
  {Fosalba}, {Frailis}, {Fraisse}, {Franceschi}, {Freschi}, {Fromenteau},
  {Frommert}, {Gaier}, {Galeotta}, {Gallegos}, {Galli}, {Gandolfo}, {Ganga},
  {Gauthier}, {G{\'e}nova-Santos}, {Ghosh}, {Giard}, {Giardino}, {Gilfanov},
  {Girard}, {Giraud-H{\'e}raud}, {Gjerl{\o}w}, {Gonz{\'a}lez-Nuevo},
  {G{\'o}rski}, {Gratton}, {Gregorio}, {Gruppuso}, {Gudmundsson}, {Haissinski},
  {Hamann}, {Hansen}, {Hansen}, {Hanson}, {Harrison}, {Heavens}, {Helou},
  {Hempel}, {Henrot-Versill{\'e}}, {Hern{\'a}ndez-Monteagudo}, {Herranz},
  {Hildebrandt}, {Hivon}, {Ho}, {Hobson}, {Holmes}, {Hornstrup}, {Hou},
  {Hovest}, {Huey}, {Huffenberger}, {Hurier}, {Ili{\'c}}, {Jaffe}, {Jaffe},
  {Jasche}, {Jewell}, {Jones}, {Juvela}, {Kalberla}, {Kangaslahti},
  {Keih{\"a}nen}, {Kerp}, {Keskitalo}, {Khamitov}, {Kiiveri}, {Kim}, {Kisner},
  {Kneissl}, {Knoche}, {Knox}, {Kunz}, {Kurki-Suonio}, {Lacasa}, {Lagache},
  {L{\"a}hteenm{\"a}ki}, {Lamarre}, {Langer}, {Lasenby}, {Lattanzi},
  {Laureijs}, {Lavabre}, {Lawrence}, {Le Jeune}, {Leach}, {Leahy}, {Leonardi},
  {Le{\'o}n-Tavares}, {Leroy}, {Lesgourgues}, {Lewis}, {Li}, {Liddle},
  {Liguori}, {Lilje}, {Linden-V{\o}rnle}, {Lindholm}, {L{\'o}pez-Caniego},
  {Lowe}, {Lubin}, {Mac{\'\i}as-P{\'e}rez}, {MacTavish}, {Maffei}, {Maggio},
  {Maino}, {Mandolesi}, {Mangilli}, {Marcos-Caballero}, {Marinucci}, {Maris},
  {Marleau}, {Marshall}, {Martin}, {Mart{\'\i}nez-Gonz{\'a}lez}, {Masi},
  {Massardi}, {Matarrese}, {Matsumura}, {Matthai}, {Maurin}, {Mazzotta},
  {McDonald}, {McEwen}, {McGehee}, {Mei}, {Meinhold}, {Melchiorri}, {Melin},
  {Mendes}, {Menegoni}, {Mennella}, {Migliaccio}, {Mikkelsen}, {Millea},
  {Miniscalco}, {Mitra}, {Miville-Desch{\^e}nes}, {Molinari}, {Moneti},
  {Montier}, {Morgante}, {Morisset}, {Mortlock}, {Moss}, {Munshi}, {Murphy},
  {Naselsky}, {Nati}, {Natoli}, {Negrello}, {Nesvadba}, {Netterfield},
  {N{\o}rgaard-Nielsen}, {North}, {Noviello}, {Novikov}, {Novikov}, {O'Dwyer},
  {Orieux}, {Osborne}, {O'Sullivan}, {Oxborrow}, {Paci}, {Pagano}, {Pajot},
  {Paladini}, {Pandolfi}, {Paoletti}, {Partridge}, {Pasian}, {Patanchon},
  {Paykari}, {Pearson}, {Pearson}, {Peel}, {Peiris}, {Perdereau}, {Perotto},
  {Perrotta}, {Pettorino}, {Piacentini}, {Piat}, {Pierpaoli}, {Pietrobon},
  {Plaszczynski}, {Platania}, {Pogosyan}, {Pointecouteau}, {Polenta},
  {Ponthieu}, {Popa}, {Poutanen}, {Pratt}, {Pr{\'e}zeau}, {Prunet}, {Puget},
  {Pullen}, {Rachen}, {Racine}, {Rahlin}, {R{\"a}th}, {Reach}, {Rebolo},
  {Reinecke}, {Remazeilles}, {Renault}, {Renzi}, {Riazuelo}, {Ricciardi},
  {Riller}, {Ringeval}, {Ristorcelli}, {Robbers}, {Rocha}, {Roman}, {Rosset},
  {Rossetti}, {Roudier}, {Rowan-Robinson}, {Rubi{\~n}o-Mart{\'\i}n},
  {Ruiz-Granados}, {Rusholme}, {Salerno}, {Sandri}, {Sanselme}, {Santos},
  {Savelainen}, {Savini}, {Schaefer}, {Schiavon}, {Scott}, {Seiffert}, {Serra},
  {Shellard}, {Smith}, {Smoot}, {Souradeep}, {Spencer}, {Starck}, {Stolyarov},
  {Stompor}, {Sudiwala}, {Sunyaev}, {Sureau}, {Sutter}, {Sutton}, {Suur-Uski},
  {Sygnet}, {Tauber}, {Tavagnacco}, {Taylor}, {Terenzi}, {Texier},
  {Toffolatti}, {Tomasi}, {Torre}, {Tristram}, {Tucci}, {Tuovinen},
  {T{\"u}rler}, {Tuttlebee}, {Umana}, {Valenziano}, {Valiviita}, {Van Tent},
  {Varis}, {Vibert}, {Viel}, {Vielva}, {Villa}, {Vittorio}, {Wade}, {Wandelt},
  {Watson}, {Watson}, {Wehus}, {Welikala}, {Weller}, {White}, {White},
  {Wilkinson}, {Winkel}, {Xia}, {Yvon}, {Zacchei}, {Zibin}, \&
  {Zonca}}]{2014A&A...571A...1P}
{Planck Collaboration}, {Ade}, P.~A.~R., {Aghanim}, N., {et~al.}
  2014{\natexlab{b}}, \aap, 571, A1

\bibitem[{{Planck Collaboration} {et~al.}(2015){Planck Collaboration}, {Ade},
  {Alves}, {Aniano}, {Armitage-Caplan}, {Arnaud}, {Atrio-Barandela}, {Aumont},
  {Baccigalupi}, {Banday}, {Barreiro}, {Battaner}, {Benabed},
  {Benoit-L{\'e}vy}, {Bernard}, {Bersanelli}, {Bielewicz}, {Bock}, {Bond},
  {Borrill}, {Bouchet}, {Boulanger}, {Burigana}, {Cardoso}, {Catalano},
  {Chamballu}, {Chiang}, {Colombo}, {Combet}, {Couchot}, {Coulais}, {Crill},
  {Curto}, {Cuttaia}, {Danese}, {Davies}, {Davis}, {de Bernardis}, {de Zotti},
  {Delabrouille}, {D{\'e}sert}, {Dickinson}, {Diego}, {Donzelli}, {Dor{\'e}},
  {Douspis}, {Dunkley}, {Dupac}, {En{\ss}lin}, {Eriksen}, {Falgarone},
  {Finelli}, {Forni}, {Frailis}, {Fraisse}, {Franceschi}, {Galeotta}, {Ganga},
  {Ghosh}, {Giard}, {Gonz{\'a}lez-Nuevo}, {G{\'o}rski}, {Gregorio}, {Gruppuso},
  {Guillet}, {Hansen}, {Harrison}, {Helou}, {Hern{\'a}ndez-Monteagudo},
  {Hildebrandt}, {Hivon}, {Hobson}, {Holmes}, {Hornstrup}, {Jaffe}, {Jaffe},
  {Jones}, {Keih{\"a}nen}, {Keskitalo}, {Kisner}, {Kneissl}, {Knoche}, {Kunz},
  {Kurki-Suonio}, {Lagache}, {Lamarre}, {Lasenby}, {Lawrence}, {Leahy},
  {Leonardi}, {Levrier}, {Liguori}, {Lilje}, {Linden-V{\o}rnle},
  {L{\'o}pez-Caniego}, {Lubin}, {Mac{\'\i}as-P{\'e}rez}, {Maffei},
  {Magalh{\~a}es}, {Maino}, {Mandolesi}, {Maris}, {Marshall}, {Martin},
  {Mart{\'\i}nez-Gonz{\'a}lez}, {Masi}, {Matarrese}, {Mazzotta}, {Melchiorri},
  {Mendes}, {Mennella}, {Migliaccio}, {Miville-Desch{\^e}nes}, {Moneti},
  {Montier}, {Morgante}, {Mortlock}, {Munshi}, {Murphy}, {Naselsky}, {Nati},
  {Natoli}, {Netterfield}, {Noviello}, {Novikov}, {Novikov}, {Oppermann},
  {Oxborrow}, {Pagano}, {Pajot}, {Paoletti}, {Pasian}, {Perdereau}, {Perotto},
  {Perrotta}, {Piacentini}, {Pietrobon}, {Plaszczynski}, {Pointecouteau},
  {Polenta}, {Popa}, {Pratt}, {Rachen}, {Reach}, {Reinecke}, {Remazeilles},
  {Renault}, {Ricciardi}, {Riller}, {Ristorcelli}, {Rocha}, {Rosset},
  {Roudier}, {Rubi{\~n}o-Mart{\'\i}n}, {Rusholme}, {Salerno}, {Sandri},
  {Savini}, {Scott}, {Spencer}, {Stolyarov}, {Stompor}, {Sudiwala}, {Sutton},
  {Suur-Uski}, {Sygnet}, {Tauber}, {Terenzi}, {Toffolatti}, {Tomasi},
  {Tristram}, {Tucci}, {Valenziano}, {Valiviita}, {Van Tent}, {Vielva},
  {Villa}, {Wandelt}, {Zacchei}, \& {Zonca}}]{2015A&A...576A.107P}
{Planck Collaboration}, {Ade}, P.~A.~R., {Alves}, M.~I.~R., {et~al.} 2015,
  \aap, 576, A107

\bibitem[{{Planck Collaboration} {et~al.}(2020{\natexlab{a}}){Planck
  Collaboration}, {Aghanim}, {Akrami}, {Alves}, {Ashdown}, {Aumont},
  {Baccigalupi}, {Ballardini}, {Banday}, {Barreiro}, {Bartolo}, {Basak},
  {Benabed}, {Bernard}, {Bersanelli}, {Bielewicz}, {Bock}, {Bond}, {Borrill},
  {Bouchet}, {Boulanger}, {Bracco}, {Bucher}, {Burigana}, {Calabrese},
  {Cardoso}, {Carron}, {Chary}, {Chiang}, {Colombo}, {Combet}, {Crill},
  {Cuttaia}, {de Bernardis}, {de Zotti}, {Delabrouille}, {Delouis}, {Di
  Valentino}, {Dickinson}, {Diego}, {Dor{\'e}}, {Douspis}, {Ducout}, {Dupac},
  {Efstathiou}, {Elsner}, {En{\ss}lin}, {Eriksen}, {Falgarone}, {Fantaye},
  {Fernandez-Cobos}, {Ferri{\`e}re}, {Finelli}, {Forastieri}, {Frailis},
  {Fraisse}, {Franceschi}, {Frolov}, {Galeotta}, {Galli}, {Ganga},
  {G{\'e}nova-Santos}, {Gerbino}, {Ghosh}, {Gonz{\'a}lez-Nuevo}, {G{\'o}rski},
  {Gratton}, {Green}, {Gruppuso}, {Gudmundsson}, {Guillet}, {Handley},
  {Hansen}, {Helou}, {Herranz}, {Hivon}, {Huang}, {Jaffe}, {Jones},
  {Keih{\"a}nen}, {Keskitalo}, {Kiiveri}, {Kim}, {Krachmalnicoff}, {Kunz},
  {Kurki-Suonio}, {Lagache}, {Lamarre}, {Lasenby}, {Lattanzi}, {Lawrence}, {Le
  Jeune}, {Levrier}, {Liguori}, {Lilje}, {Lindholm}, {L{\'o}pez-Caniego},
  {Lubin}, {Ma}, {Mac{\'\i}as-P{\'e}rez}, {Maggio}, {Maino}, {Mandolesi},
  {Mangilli}, {Marcos-Caballero}, {Maris}, {Martin},
  {Mart{\'\i}nez-Gonz{\'a}lez}, {Matarrese}, {Mauri}, {McEwen}, {Melchiorri},
  {Mennella}, {Migliaccio}, {Miville-Desch{\^e}nes}, {Molinari}, {Moneti},
  {Montier}, {Morgante}, {Moss}, {Natoli}, {Pagano}, {Paoletti}, {Patanchon},
  {Perrotta}, {Pettorino}, {Piacentini}, {Polastri}, {Polenta}, {Puget},
  {Rachen}, {Reinecke}, {Remazeilles}, {Renzi}, {Ristorcelli}, {Rocha},
  {Rosset}, {Roudier}, {Rubi{\~n}o-Mart{\'\i}n}, {Ruiz-Granados}, {Salvati},
  {Sandri}, {Savelainen}, {Scott}, {Sirignano}, {Sunyaev}, {Suur-Uski},
  {Tauber}, {Tavagnacco}, {Tenti}, {Toffolatti}, {Tomasi}, {Trombetti},
  {Valiviita}, {Vansyngel}, {Van Tent}, {Vielva}, {Villa}, {Vittorio},
  {Wandelt}, {Wehus}, {Zacchei}, \& {Zonca}}]{2020A&A...641A..12P}
{Planck Collaboration}, {Aghanim}, N., {Akrami}, Y., {et~al.}
  2020{\natexlab{a}}, \aap, 641, A12

\bibitem[{{Planck Collaboration} {et~al.}(2020{\natexlab{b}}){Planck
  Collaboration}, {Aghanim}, {Akrami}, {Arroja}, {Ashdown}, {Aumont},
  {Baccigalupi}, {Ballardini}, {Banday}, {Barreiro}, {Bartolo}, {Basak},
  {Battye}, {Benabed}, {Bernard}, {Bersanelli}, {Bielewicz}, {Bock}, {Bond},
  {Borrill}, {Bouchet}, {Boulanger}, {Bucher}, {Burigana}, {Butler},
  {Calabrese}, {Cardoso}, {Carron}, {Casaponsa}, {Challinor}, {Chiang},
  {Colombo}, {Combet}, {Contreras}, {Crill}, {Cuttaia}, {de Bernardis}, {de
  Zotti}, {Delabrouille}, {Delouis}, {D{\'e}sert}, {Di Valentino}, {Dickinson},
  {Diego}, {Donzelli}, {Dor{\'e}}, {Douspis}, {Ducout}, {Dupac}, {Efstathiou},
  {Elsner}, {En{\ss}lin}, {Eriksen}, {Falgarone}, {Fantaye}, {Fergusson},
  {Fernandez-Cobos}, {Finelli}, {Forastieri}, {Frailis}, {Franceschi},
  {Frolov}, {Galeotta}, {Galli}, {Ganga}, {G{\'e}nova-Santos}, {Gerbino},
  {Ghosh}, {Gonz{\'a}lez-Nuevo}, {G{\'o}rski}, {Gratton}, {Gruppuso},
  {Gudmundsson}, {Hamann}, {Handley}, {Hansen}, {Helou}, {Herranz},
  {Hildebrandt}, {Hivon}, {Huang}, {Jaffe}, {Jones}, {Karakci}, {Keih{\"a}nen},
  {Keskitalo}, {Kiiveri}, {Kim}, {Kisner}, {Knox}, {Krachmalnicoff}, {Kunz},
  {Kurki-Suonio}, {Lagache}, {Lamarre}, {Langer}, {Lasenby}, {Lattanzi},
  {Lawrence}, {Le Jeune}, {Leahy}, {Lesgourgues}, {Levrier}, {Lewis},
  {Liguori}, {Lilje}, {Lilley}, {Lindholm}, {L{\'o}pez-Caniego}, {Lubin}, {Ma},
  {Mac{\'\i}as-P{\'e}rez}, {Maggio}, {Maino}, {Mandolesi}, {Mangilli},
  {Marcos-Caballero}, {Maris}, {Martin}, {Martinelli},
  {Mart{\'\i}nez-Gonz{\'a}lez}, {Matarrese}, {Mauri}, {McEwen}, {Meerburg},
  {Meinhold}, {Melchiorri}, {Mennella}, {Migliaccio}, {Millea}, {Mitra},
  {Miville-Desch{\^e}nes}, {Molinari}, {Moneti}, {Montier}, {Morgante}, {Moss},
  {Mottet}, {M{\"u}nchmeyer}, {Natoli}, {N{\o}rgaard-Nielsen}, {Oxborrow},
  {Pagano}, {Paoletti}, {Partridge}, {Patanchon}, {Pearson}, {Peel}, {Peiris},
  {Perrotta}, {Pettorino}, {Piacentini}, {Polastri}, {Polenta}, {Puget},
  {Rachen}, {Reinecke}, {Remazeilles}, {Renault}, {Renzi}, {Rocha}, {Rosset},
  {Roudier}, {Rubi{\~n}o-Mart{\'\i}n}, {Ruiz-Granados}, {Salvati}, {Sandri},
  {Savelainen}, {Scott}, {Shellard}, {Shiraishi}, {Sirignano}, {Sirri},
  {Spencer}, {Sunyaev}, {Suur-Uski}, {Tauber}, {Tavagnacco}, {Tenti},
  {Terenzi}, {Toffolatti}, {Tomasi}, {Trombetti}, {Valiviita}, {Van Tent},
  {Vibert}, {Vielva}, {Villa}, {Vittorio}, {Wandelt}, {Wehus}, {White},
  {White}, {Zacchei}, \& {Zonca}}]{2020A&A...641A...1P}
{Planck Collaboration}, {Aghanim}, N., {Akrami}, Y., {et~al.}
  2020{\natexlab{b}}, \aap, 641, A1

\bibitem[{{Planck Collaboration} {et~al.}(2020{\natexlab{c}}){Planck
  Collaboration}, {Aghanim}, {Akrami}, {Ashdown}, {Aumont}, {Baccigalupi},
  {Ballardini}, {Banday}, {Barreiro}, {Bartolo}, {Basak}, {Battye}, {Benabed},
  {Bernard}, {Bersanelli}, {Bielewicz}, {Bock}, {Bond}, {Borrill}, {Bouchet},
  {Boulanger}, {Bucher}, {Burigana}, {Butler}, {Calabrese}, {Cardoso},
  {Carron}, {Challinor}, {Chiang}, {Chluba}, {Colombo}, {Combet}, {Contreras},
  {Crill}, {Cuttaia}, {de Bernardis}, {de Zotti}, {Delabrouille}, {Delouis},
  {Di Valentino}, {Diego}, {Dor{\'e}}, {Douspis}, {Ducout}, {Dupac}, {Dusini},
  {Efstathiou}, {Elsner}, {En{\ss}lin}, {Eriksen}, {Fantaye}, {Farhang},
  {Fergusson}, {Fernandez-Cobos}, {Finelli}, {Forastieri}, {Frailis},
  {Fraisse}, {Franceschi}, {Frolov}, {Galeotta}, {Galli}, {Ganga},
  {G{\'e}nova-Santos}, {Gerbino}, {Ghosh}, {Gonz{\'a}lez-Nuevo}, {G{\'o}rski},
  {Gratton}, {Gruppuso}, {Gudmundsson}, {Hamann}, {Handley}, {Hansen},
  {Herranz}, {Hildebrandt}, {Hivon}, {Huang}, {Jaffe}, {Jones}, {Karakci},
  {Keih{\"a}nen}, {Keskitalo}, {Kiiveri}, {Kim}, {Kisner}, {Knox},
  {Krachmalnicoff}, {Kunz}, {Kurki-Suonio}, {Lagache}, {Lamarre}, {Lasenby},
  {Lattanzi}, {Lawrence}, {Le Jeune}, {Lemos}, {Lesgourgues}, {Levrier},
  {Lewis}, {Liguori}, {Lilje}, {Lilley}, {Lindholm}, {L{\'o}pez-Caniego},
  {Lubin}, {Ma}, {Mac{\'\i}as-P{\'e}rez}, {Maggio}, {Maino}, {Mandolesi},
  {Mangilli}, {Marcos-Caballero}, {Maris}, {Martin}, {Martinelli},
  {Mart{\'\i}nez-Gonz{\'a}lez}, {Matarrese}, {Mauri}, {McEwen}, {Meinhold},
  {Melchiorri}, {Mennella}, {Migliaccio}, {Millea}, {Mitra},
  {Miville-Desch{\^e}nes}, {Molinari}, {Montier}, {Morgante}, {Moss}, {Natoli},
  {N{\o}rgaard-Nielsen}, {Pagano}, {Paoletti}, {Partridge}, {Patanchon},
  {Peiris}, {Perrotta}, {Pettorino}, {Piacentini}, {Polastri}, {Polenta},
  {Puget}, {Rachen}, {Reinecke}, {Remazeilles}, {Renzi}, {Rocha}, {Rosset},
  {Roudier}, {Rubi{\~n}o-Mart{\'\i}n}, {Ruiz-Granados}, {Salvati}, {Sandri},
  {Savelainen}, {Scott}, {Shellard}, {Sirignano}, {Sirri}, {Spencer},
  {Sunyaev}, {Suur-Uski}, {Tauber}, {Tavagnacco}, {Tenti}, {Toffolatti},
  {Tomasi}, {Trombetti}, {Valenziano}, {Valiviita}, {Van Tent}, {Vibert},
  {Vielva}, {Villa}, {Vittorio}, {Wandelt}, {Wehus}, {White}, {White},
  {Zacchei}, \& {Zonca}}]{2020A&A...641A...6P}
{Planck Collaboration}, {Aghanim}, N., {Akrami}, Y., {et~al.}
  2020{\natexlab{c}}, \aap, 641, A6

\bibitem[{{Planck Collaboration} {et~al.}(2020{\natexlab{d}}){Planck
  Collaboration}, {Aghanim}, {Akrami}, {Ashdown}, {Aumont}, {Baccigalupi},
  {Ballardini}, {Banday}, {Barreiro}, {Bartolo}, {Basak}, {Benabed}, {Bernard},
  {Bersanelli}, {Bielewicz}, {Bond}, {Borrill}, {Bouchet}, {Boulanger},
  {Bucher}, {Burigana}, {Calabrese}, {Cardoso}, {Carron}, {Challinor},
  {Chiang}, {Colombo}, {Combet}, {Couchot}, {Crill}, {Cuttaia}, {de Bernardis},
  {de Rosa}, {de Zotti}, {Delabrouille}, {Delouis}, {Di Valentino}, {Diego},
  {Dor{\'e}}, {Douspis}, {Ducout}, {Dupac}, {Efstathiou}, {Elsner},
  {En{\ss}lin}, {Eriksen}, {Falgarone}, {Fantaye}, {Finelli}, {Frailis},
  {Fraisse}, {Franceschi}, {Frolov}, {Galeotta}, {Galli}, {Ganga},
  {G{\'e}nova-Santos}, {Gerbino}, {Ghosh}, {Gonz{\'a}lez-Nuevo}, {G{\'o}rski},
  {Gratton}, {Gruppuso}, {Gudmundsson}, {Handley}, {Hansen},
  {Henrot-Versill{\'e}}, {Herranz}, {Hivon}, {Huang}, {Jaffe}, {Jones},
  {Karakci}, {Keih{\"a}nen}, {Keskitalo}, {Kiiveri}, {Kim}, {Kisner},
  {Krachmalnicoff}, {Kunz}, {Kurki-Suonio}, {Lagache}, {Lamarre}, {Lasenby},
  {Lattanzi}, {Lawrence}, {Levrier}, {Liguori}, {Lilje}, {Lindholm},
  {L{\'o}pez-Caniego}, {Ma}, {Mac{\'\i}as-P{\'e}rez}, {Maggio}, {Maino},
  {Mandolesi}, {Mangilli}, {Martin}, {Mart{\'\i}nez-Gonz{\'a}lez}, {Matarrese},
  {Mauri}, {McEwen}, {Melchiorri}, {Mennella}, {Migliaccio},
  {Miville-Desch{\^e}nes}, {Molinari}, {Moneti}, {Montier}, {Morgante}, {Moss},
  {Mottet}, {Natoli}, {Pagano}, {Paoletti}, {Partridge}, {Patanchon},
  {Patrizii}, {Perdereau}, {Perrotta}, {Pettorino}, {Piacentini}, {Puget},
  {Rachen}, {Reinecke}, {Remazeilles}, {Renzi}, {Rocha}, {Roudier}, {Salvati},
  {Sandri}, {Savelainen}, {Scott}, {Sirignano}, {Sirri}, {Spencer}, {Sunyaev},
  {Suur-Uski}, {Tauber}, {Tavagnacco}, {Tenti}, {Toffolatti}, {Tomasi},
  {Tristram}, {Trombetti}, {Valiviita}, {Vansyngel}, {Van Tent}, {Vibert},
  {Vielva}, {Villa}, {Vittorio}, {Wandelt}, {Wehus}, \&
  {Zonca}}]{2020A&A...641A...3P}
{Planck Collaboration}, {Aghanim}, N., {Akrami}, Y., {et~al.}
  2020{\natexlab{d}}, \aap, 641, A3

\bibitem[{{Planck Collaboration} {et~al.}(2020{\natexlab{e}}){Planck
  Collaboration}, {Akrami}, {Arg{\"u}eso}, {Ashdown}, {Aumont}, {Baccigalupi},
  {Ballardini}, {Banday}, {Barreiro}, {Bartolo}, {Basak}, {Benabed}, {Bernard},
  {Bersanelli}, {Bielewicz}, {Bonavera}, {Bond}, {Borrill}, {Bouchet},
  {Boulanger}, {Bucher}, {Burigana}, {Butler}, {Calabrese}, {Cardoso},
  {Colombo}, {Crill}, {Cuttaia}, {de Bernardis}, {de Rosa}, {de Zotti},
  {Delabrouille}, {Di Valentino}, {Dickinson}, {Diego}, {Donzelli}, {Ducout},
  {Dupac}, {Efstathiou}, {Elsner}, {En{\ss}lin}, {Eriksen}, {Fantaye},
  {Finelli}, {Frailis}, {Franceschi}, {Frolov}, {Galeotta}, {Galli}, {Ganga},
  {G{\'e}nova-Santos}, {Gerbino}, {Ghosh}, {Gonz{\'a}lez-Nuevo}, {G{\'o}rski},
  {Gratton}, {Gruppuso}, {Gudmundsson}, {Handley}, {Hansen}, {Herranz},
  {Hivon}, {Huang}, {Jaffe}, {Jones}, {Karakci}, {Keih{\"a}nen}, {Keskitalo},
  {Kiiveri}, {Kim}, {Kisner}, {Krachmalnicoff}, {Kunz}, {Kurki-Suonio},
  {Lamarre}, {Lasenby}, {Lattanzi}, {Lawrence}, {Leahy}, {Levrier}, {Liguori},
  {Lilje}, {Lindholm}, {L{\'o}pez-Caniego}, {Ma}, {Mac{\'\i}as-P{\'e}rez},
  {Maggio}, {Maino}, {Mandolesi}, {Mangilli}, {Maris}, {Martin},
  {Mart{\'\i}nez-Gonz{\'a}lez}, {Matarrese}, {Mauri}, {McEwen}, {Meinhold},
  {Melchiorri}, {Mennella}, {Migliaccio}, {Molinari}, {Montier}, {Morgante},
  {Moss}, {Natoli}, {Pagano}, {Paoletti}, {Partridge}, {Patanchon}, {Patrizii},
  {Peel}, {Perrotta}, {Pettorino}, {Piacentini}, {Polenta}, {Puget}, {Rachen},
  {Racine}, {Reinecke}, {Remazeilles}, {Renzi}, {Rocha}, {Roudier},
  {Rubi{\~n}o-Mart{\'\i}n}, {Salvati}, {Sandri}, {Savelainen}, {Scott},
  {Seljebotn}, {Sirignano}, {Sirri}, {Spencer}, {Suur-Uski}, {Tauber},
  {Tavagnacco}, {Tenti}, {Terenzi}, {Toffolatti}, {Tomasi}, {Trombetti},
  {Valiviita}, {Vansyngel}, {Van Tent}, {Vielva}, {Villa}, {Vittorio},
  {Wandelt}, {Watson}, {Wehus}, {Zacchei}, \& {Zonca}}]{2020A&A...641A...2P}
{Planck Collaboration}, {Akrami}, Y., {Arg{\"u}eso}, F., {et~al.}
  2020{\natexlab{e}}, \aap, 641, A2

\bibitem[{{Planck Collaboration} {et~al.}(2020{\natexlab{f}}){Planck
  Collaboration}, {Akrami}, {Arroja}, {Ashdown}, {Aumont}, {Baccigalupi},
  {Ballardini}, {Banday}, {Barreiro}, {Bartolo}, {Basak}, {Benabed}, {Bernard},
  {Bersanelli}, {Bielewicz}, {Bock}, {Bond}, {Borrill}, {Bouchet}, {Boulanger},
  {Bucher}, {Burigana}, {Butler}, {Calabrese}, {Cardoso}, {Carron},
  {Challinor}, {Chiang}, {Colombo}, {Combet}, {Contreras}, {Crill}, {Cuttaia},
  {de Bernardis}, {de Zotti}, {Delabrouille}, {Delouis}, {Di Valentino},
  {Diego}, {Donzelli}, {Dor{\'e}}, {Douspis}, {Ducout}, {Dupac}, {Dusini},
  {Efstathiou}, {Elsner}, {En{\ss}lin}, {Eriksen}, {Fantaye}, {Fergusson},
  {Fernandez-Cobos}, {Finelli}, {Forastieri}, {Frailis}, {Franceschi},
  {Frolov}, {Galeotta}, {Galli}, {Ganga}, {Gauthier}, {G{\'e}nova-Santos},
  {Gerbino}, {Ghosh}, {Gonz{\'a}lez-Nuevo}, {G{\'o}rski}, {Gratton},
  {Gruppuso}, {Gudmundsson}, {Hamann}, {Handley}, {Hansen}, {Herranz}, {Hivon},
  {Hooper}, {Huang}, {Jaffe}, {Jones}, {Keih{\"a}nen}, {Keskitalo}, {Kiiveri},
  {Kim}, {Kisner}, {Krachmalnicoff}, {Kunz}, {Kurki-Suonio}, {Lagache},
  {Lamarre}, {Lasenby}, {Lattanzi}, {Lawrence}, {Le Jeune}, {Lesgourgues},
  {Levrier}, {Lewis}, {Liguori}, {Lilje}, {Lindholm}, {L{\'o}pez-Caniego},
  {Lubin}, {Ma}, {Mac{\'\i}as-P{\'e}rez}, {Maggio}, {Maino}, {Mandolesi},
  {Mangilli}, {Marcos-Caballero}, {Maris}, {Martin},
  {Mart{\'\i}nez-Gonz{\'a}lez}, {Matarrese}, {Mauri}, {McEwen}, {Meerburg},
  {Meinhold}, {Melchiorri}, {Mennella}, {Migliaccio}, {Mitra},
  {Miville-Desch{\^e}nes}, {Molinari}, {Moneti}, {Montier}, {Morgante}, {Moss},
  {M{\"u}nchmeyer}, {Natoli}, {N{\o}rgaard-Nielsen}, {Pagano}, {Paoletti},
  {Partridge}, {Patanchon}, {Peiris}, {Perrotta}, {Pettorino}, {Piacentini},
  {Polastri}, {Polenta}, {Puget}, {Rachen}, {Reinecke}, {Remazeilles}, {Renzi},
  {Rocha}, {Rosset}, {Roudier}, {Rubi{\~n}o-Mart{\'\i}n}, {Ruiz-Granados},
  {Salvati}, {Sandri}, {Savelainen}, {Scott}, {Shellard}, {Shiraishi},
  {Sirignano}, {Sirri}, {Spencer}, {Sunyaev}, {Suur-Uski}, {Tauber},
  {Tavagnacco}, {Tenti}, {Toffolatti}, {Tomasi}, {Trombetti}, {Valiviita}, {Van
  Tent}, {Vielva}, {Villa}, {Vittorio}, {Wandelt}, {Wehus}, {White}, {Zacchei},
  {Zibin}, \& {Zonca}}]{2020A&A...641A..10P}
{Planck Collaboration}, {Akrami}, Y., {Arroja}, F., {et~al.}
  2020{\natexlab{f}}, \aap, 641, A10

\bibitem[{{Planck Collaboration} {et~al.}(2020{\natexlab{g}}){Planck
  Collaboration}, {Akrami}, {Arroja}, {Ashdown}, {Aumont}, {Baccigalupi},
  {Ballardini}, {Banday}, {Barreiro}, {Bartolo}, {Basak}, {Benabed}, {Bernard},
  {Bersanelli}, {Bielewicz}, {Bond}, {Borrill}, {Bouchet}, {Bucher},
  {Burigana}, {Butler}, {Calabrese}, {Cardoso}, {Casaponsa}, {Challinor},
  {Chiang}, {Colombo}, {Combet}, {Crill}, {Cuttaia}, {de Bernardis}, {de Rosa},
  {de Zotti}, {Delabrouille}, {Delouis}, {Di Valentino}, {Diego}, {Dor{\'e}},
  {Douspis}, {Ducout}, {Dupac}, {Dusini}, {Efstathiou}, {Elsner}, {En{\ss}lin},
  {Eriksen}, {Fantaye}, {Fergusson}, {Fernandez-Cobos}, {Finelli}, {Frailis},
  {Fraisse}, {Franceschi}, {Frolov}, {Galeotta}, {Galli}, {Ganga},
  {G{\'e}nova-Santos}, {Gerbino}, {Gonz{\'a}lez-Nuevo}, {G{\'o}rski},
  {Gratton}, {Gruppuso}, {Gudmundsson}, {Hamann}, {Handley}, {Hansen},
  {Herranz}, {Hivon}, {Huang}, {Jaffe}, {Jones}, {Jung}, {Keih{\"a}nen},
  {Keskitalo}, {Kiiveri}, {Kim}, {Krachmalnicoff}, {Kunz}, {Kurki-Suonio},
  {Lamarre}, {Lasenby}, {Lattanzi}, {Lawrence}, {Le Jeune}, {Levrier}, {Lewis},
  {Liguori}, {Lilje}, {Lindholm}, {L{\'o}pez-Caniego}, {Ma},
  {Mac{\'\i}as-P{\'e}rez}, {Maggio}, {Maino}, {Mandolesi}, {Marcos-Caballero},
  {Maris}, {Martin}, {Mart{\'\i}nez-Gonz{\'a}lez}, {Matarrese}, {Mauri},
  {McEwen}, {Meerburg}, {Meinhold}, {Melchiorri}, {Mennella}, {Migliaccio},
  {Miville-Desch{\^e}nes}, {Molinari}, {Moneti}, {Montier}, {Morgante}, {Moss},
  {M{\"u}nchmeyer}, {Natoli}, {Oppizzi}, {Pagano}, {Paoletti}, {Partridge},
  {Patanchon}, {Perrotta}, {Pettorino}, {Piacentini}, {Polenta}, {Puget},
  {Rachen}, {Racine}, {Reinecke}, {Remazeilles}, {Renzi}, {Rocha},
  {Rubi{\~n}o-Mart{\'\i}n}, {Ruiz-Granados}, {Salvati}, {Savelainen}, {Scott},
  {Shellard}, {Shiraishi}, {Sirignano}, {Sirri}, {Smith}, {Spencer}, {Stanco},
  {Sunyaev}, {Suur-Uski}, {Tauber}, {Tavagnacco}, {Tenti}, {Toffolatti},
  {Tomasi}, {Trombetti}, {Valiviita}, {Van Tent}, {Vielva}, {Villa},
  {Vittorio}, {Wandelt}, {Wehus}, {Zacchei}, \& {Zonca}}]{2020A&A...641A...9P}
{Planck Collaboration}, {Akrami}, Y., {Arroja}, F., {et~al.}
  2020{\natexlab{g}}, \aap, 641, A9

\bibitem[{{Planck Collaboration} {et~al.}(2020{\natexlab{h}}){Planck
  Collaboration}, {Akrami}, {Ashdown}, {Aumont}, {Baccigalupi}, {Ballardini},
  {Banday}, {Barreiro}, {Bartolo}, {Basak}, {Benabed}, {Bernard}, {Bersanelli},
  {Bielewicz}, {Bond}, {Borrill}, {Bouchet}, {Boulanger}, {Bracco}, {Bucher},
  {Burigana}, {Calabrese}, {Cardoso}, {Carron}, {Chiang}, {Combet}, {Crill},
  {de Bernardis}, {de Zotti}, {Delabrouille}, {Delouis}, {Di Valentino},
  {Dickinson}, {Diego}, {Ducout}, {Dupac}, {Efstathiou}, {Elsner},
  {En{\ss}lin}, {Falgarone}, {Fantaye}, {Ferri{\`e}re}, {Finelli},
  {Forastieri}, {Frailis}, {Fraisse}, {Franceschi}, {Frolov}, {Galeotta},
  {Galli}, {Ganga}, {G{\'e}nova-Santos}, {Ghosh}, {Gonz{\'a}lez-Nuevo},
  {G{\'o}rski}, {Gruppuso}, {Gudmundsson}, {Guillet}, {Handley}, {Hansen},
  {Herranz}, {Huang}, {Jaffe}, {Jones}, {Keih{\"a}nen}, {Keskitalo}, {Kiiveri},
  {Kim}, {Krachmalnicoff}, {Kunz}, {Kurki-Suonio}, {Lamarre}, {Lasenby}, {Le
  Jeune}, {Levrier}, {Liguori}, {Lilje}, {Lindholm}, {L{\'o}pez-Caniego},
  {Lubin}, {Ma}, {Mac{\'\i}as-P{\'e}rez}, {Maggio}, {Maino}, {Mandolesi},
  {Mangilli}, {Martin}, {Mart{\'\i}nez-Gonz{\'a}lez}, {Matarrese}, {McEwen},
  {Meinhold}, {Melchiorri}, {Migliaccio}, {Miville-Desch{\^e}nes}, {Molinari},
  {Moneti}, {Montier}, {Morgante}, {Natoli}, {Pagano}, {Paoletti}, {Pettorino},
  {Piacentini}, {Polenta}, {Puget}, {Rachen}, {Reinecke}, {Remazeilles},
  {Renzi}, {Rocha}, {Rosset}, {Roudier}, {Rubi{\~n}o-Mart{\'\i}n},
  {Ruiz-Granados}, {Salvati}, {Sandri}, {Savelainen}, {Scott}, {Soler},
  {Spencer}, {Tauber}, {Tavagnacco}, {Toffolatti}, {Tomasi}, {Trombetti},
  {Valiviita}, {Vansyngel}, {Van Tent}, {Vielva}, {Villa}, {Vittorio}, {Wehus},
  {Zacchei}, \& {Zonca}}]{2020A&A...641A..11P}
{Planck Collaboration}, {Akrami}, Y., {Ashdown}, M., {et~al.}
  2020{\natexlab{h}}, \aap, 641, A11

\bibitem[{{Planck Collaboration} {et~al.}(2020{\natexlab{i}}){Planck
  Collaboration}, {Akrami}, {Ashdown}, {Aumont}, {Baccigalupi}, {Ballardini},
  {Banday}, {Barreiro}, {Bartolo}, {Basak}, {Benabed}, {Bersanelli},
  {Bielewicz}, {Bock}, {Bond}, {Borrill}, {Bouchet}, {Boulanger}, {Bucher},
  {Burigana}, {Butler}, {Calabrese}, {Cardoso}, {Casaponsa}, {Chiang},
  {Colombo}, {Combet}, {Contreras}, {Crill}, {de Bernardis}, {de Zotti},
  {Delabrouille}, {Delouis}, {Di Valentino}, {Diego}, {Dor{\'e}}, {Douspis},
  {Ducout}, {Dupac}, {Efstathiou}, {Elsner}, {En{\ss}lin}, {Eriksen},
  {Fantaye}, {Fernandez-Cobos}, {Finelli}, {Frailis}, {Fraisse}, {Franceschi},
  {Frolov}, {Galeotta}, {Galli}, {Ganga}, {G{\'e}nova-Santos}, {Gerbino},
  {Ghosh}, {Gonz{\'a}lez-Nuevo}, {G{\'o}rski}, {Gruppuso}, {Gudmundsson},
  {Hamann}, {Handley}, {Hansen}, {Herranz}, {Hivon}, {Huang}, {Jaffe}, {Jones},
  {Keih{\"a}nen}, {Keskitalo}, {Kiiveri}, {Kim}, {Krachmalnicoff}, {Kunz},
  {Kurki-Suonio}, {Lagache}, {Lamarre}, {Lasenby}, {Lattanzi}, {Lawrence}, {Le
  Jeune}, {Levrier}, {Liguori}, {Lilje}, {Lindholm}, {L{\'o}pez-Caniego}, {Ma},
  {Mac{\'\i}as-P{\'e}rez}, {Maggio}, {Maino}, {Mandolesi}, {Mangilli},
  {Marcos-Caballero}, {Maris}, {Martin}, {Mart{\'\i}nez-Gonz{\'a}lez},
  {Matarrese}, {Mauri}, {McEwen}, {Meinhold}, {Mennella}, {Migliaccio},
  {Miville-Desch{\^e}nes}, {Molinari}, {Moneti}, {Montier}, {Morgante}, {Moss},
  {Natoli}, {Pagano}, {Paoletti}, {Partridge}, {Perrotta}, {Pettorino},
  {Piacentini}, {Polenta}, {Puget}, {Rachen}, {Reinecke}, {Remazeilles},
  {Renzi}, {Rocha}, {Rosset}, {Roudier}, {Rubi{\~n}o-Mart{\'\i}n},
  {Ruiz-Granados}, {Salvati}, {Savelainen}, {Scott}, {Shellard}, {Sirignano},
  {Sunyaev}, {Suur-Uski}, {Tauber}, {Tavagnacco}, {Tenti}, {Toffolatti},
  {Tomasi}, {Trombetti}, {Valenziano}, {Valiviita}, {Van Tent}, {Vielva},
  {Villa}, {Vittorio}, {Wandelt}, {Wehus}, {Zacchei}, {Zibin}, \&
  {Zonca}}]{2020A&A...641A...7P}
{Planck Collaboration}, {Akrami}, Y., {Ashdown}, M., {et~al.}
  2020{\natexlab{i}}, \aap, 641, A7

\bibitem[{{Planck Collaboration} {et~al.}(2020{\natexlab{j}}){Planck
  Collaboration}, {Akrami}, {Ashdown}, {Aumont}, {Baccigalupi}, {Ballardini},
  {Banday}, {Barreiro}, {Bartolo}, {Basak}, {Benabed}, {Bersanelli},
  {Bielewicz}, {Bond}, {Borrill}, {Bouchet}, {Boulanger}, {Bucher}, {Burigana},
  {Calabrese}, {Cardoso}, {Carron}, {Casaponsa}, {Challinor}, {Colombo},
  {Combet}, {Crill}, {Cuttaia}, {de Bernardis}, {de Rosa}, {de Zotti},
  {Delabrouille}, {Delouis}, {Di Valentino}, {Dickinson}, {Diego}, {Donzelli},
  {Dor{\'e}}, {Ducout}, {Dupac}, {Efstathiou}, {Elsner}, {En{\ss}lin},
  {Eriksen}, {Falgarone}, {Fernandez-Cobos}, {Finelli}, {Forastieri},
  {Frailis}, {Fraisse}, {Franceschi}, {Frolov}, {Galeotta}, {Galli}, {Ganga},
  {G{\'e}nova-Santos}, {Gerbino}, {Ghosh}, {Gonz{\'a}lez-Nuevo}, {G{\'o}rski},
  {Gratton}, {Gruppuso}, {Gudmundsson}, {Handley}, {Hansen}, {Helou},
  {Herranz}, {Hildebrandt}, {Huang}, {Jaffe}, {Karakci}, {Keih{\"a}nen},
  {Keskitalo}, {Kiiveri}, {Kim}, {Kisner}, {Krachmalnicoff}, {Kunz},
  {Kurki-Suonio}, {Lagache}, {Lamarre}, {Lasenby}, {Lattanzi}, {Lawrence}, {Le
  Jeune}, {Levrier}, {Liguori}, {Lilje}, {Lindholm}, {L{\'o}pez-Caniego},
  {Lubin}, {Ma}, {Mac{\'\i}as-P{\'e}rez}, {Maggio}, {Maino}, {Mandolesi},
  {Mangilli}, {Marcos-Caballero}, {Maris}, {Martin},
  {Mart{\'\i}nez-Gonz{\'a}lez}, {Matarrese}, {Mauri}, {McEwen}, {Meinhold},
  {Melchiorri}, {Mennella}, {Migliaccio}, {Miville-Desch{\^e}nes}, {Molinari},
  {Moneti}, {Montier}, {Morgante}, {Natoli}, {Oppizzi}, {Pagano}, {Paoletti},
  {Partridge}, {Peel}, {Pettorino}, {Piacentini}, {Polenta}, {Puget}, {Rachen},
  {Reinecke}, {Remazeilles}, {Renzi}, {Rocha}, {Roudier},
  {Rubi{\~n}o-Mart{\'\i}n}, {Ruiz-Granados}, {Salvati}, {Sandri}, {Savelainen},
  {Scott}, {Seljebotn}, {Sirignano}, {Spencer}, {Suur-Uski}, {Tauber},
  {Tavagnacco}, {Tenti}, {Thommesen}, {Toffolatti}, {Tomasi}, {Trombetti},
  {Valiviita}, {Van Tent}, {Vielva}, {Villa}, {Vittorio}, {Wandelt}, {Wehus},
  {Zacchei}, \& {Zonca}}]{2020A&A...641A...4P}
{Planck Collaboration}, {Akrami}, Y., {Ashdown}, M., {et~al.}
  2020{\natexlab{j}}, \aap, 641, A4

\bibitem[{{Remazeilles} {et~al.}(2011){Remazeilles}, {Delabrouille}, \&
  {Cardoso}}]{2011MNRAS.418..467R}
{Remazeilles}, M., {Delabrouille}, J., \& {Cardoso}, J.-F. 2011, \mnras, 418,
  467

\bibitem[{{Schmalzing} \& {Gorski}(1998)}]{1998MNRAS.297..355S}
{Schmalzing}, J. \& {Gorski}, K.~M. 1998, \mnras, 297, 355

\bibitem[{Tomita(1990)}]{Tomita:1990}
Tomita, H. 1990, in Formation, Dynamics and Statistics of Patterns, ed.
  K.~Kawasaki, M.~Suzuki, \& A.~Onuki (World Scientific), 113--157

\bibitem[{{Wandelt} {et~al.}(2004){Wandelt}, {Larson}, \&
  {Lakshminarayanan}}]{2004PhRvD..70h3511W}
{Wandelt}, B.~D., {Larson}, D.~L., \& {Lakshminarayanan}, A. 2004, \prd, 70,
  083511

\bibitem[{{White} \& {Srednicki}(1995)}]{1995ApJ...443....6W}
{White}, M. \& {Srednicki}, M. 1995, \apj, 443, 6

\bibitem[{White {et~al.}(1994)White, Scott, \& Silk}]{White:1994sx}
White, M.~J., Scott, D., \& Silk, J. 1994, Ann. Rev. Astron. Astrophys., 32,
  319

\bibitem[{{Wolz} {et~al.}(2023){Wolz}, {Azzoni}, {Hervias-Caimapo}, {Errard},
  {Krachmalnicoff}, {Alonso}, {Baccigalupi}, {Baleato Lizancos}, {Brown},
  {Calabrese}, {Chluba}, {Dunkley}, {Fabbian}, {Galitzki}, {Jost}, {Morshed},
  \& {Nati}}]{2023arXiv230204276W}
{Wolz}, K., {Azzoni}, S., {Hervias-Caimapo}, C., {et~al.} 2023, arXiv e-prints,
  arXiv:2302.04276

\bibitem[{{Zaldarriaga} \& {Seljak}(1997)}]{1997PhRvD..55.1830Z}
{Zaldarriaga}, M. \& {Seljak}, U. 1997, \prd, 55, 1830

\end{thebibliography}

\appendix

\section{Variance of feature estimators}
\label{sec:app:variance}

Input data to be processed can be modelled as a sum of the true (potentially non-Gaussian) signal and the noise contribution
\begin{equation}
d = s+n.
\end{equation}
Ideally, $n$ would be a Gaussian random field with a zero mean and known covariance, often approximated as diagonal in pixel space, or even homogeneous. To construct the feature space for evaluating the similarity between pixels in a noisy input map $d$, the feature extraction process begins by convolving $d$ with a Gaussian kernel $b$ to obtain a Gaussian-smoothed map $\tilde{s}=b*d$.

As we are dealing with maps on a two-dimensional sphere, the homogeneous and isotropic noise contribution $\delta\tilde{s}$ is completely characterized by its two-point correlation function $\xi(\theta)$. The feature space utilized in this study is build up from maps constructed as local invariants derived solely from $\tilde{s}$, $\tilde{s}_{;a}$, and $\tilde{s}_{;ab}$ (refer to Section~\ref{sec:feature}). To quantify noise contribution to non-linear feature estimators, it is necessary to understand the joint probability distribution of the perturbed maps ${\cal F}^{(i)}$ arising from Gaussian random field noise $n$ and its covariant derivatives up to the second order. These can be organized into a vector as follows
\begin{equation}
  \vec{n}^T = \left[ n, n_{;1}, n_{;2}, n_{;11}, n_{;22}, n_{;12} \right] .
\end{equation}
The joint probability distribution function of the noise and its derivatives is also Gaussian and can be expressed as
\begin{equation}\label{A.3}
  P(\vec{n}) = \frac{1}{(2\pi)^3} \frac{1}{\sqrt{\det\Sigma}}\,
    \exp\left[ - \frac{1}{2} \vec{n}^T\Sigma^{-1}\vec{n} \right] .
\end{equation}
Here, $\Sigma$ represents the corresponding covariance matrix. A seminal study of Gaussian random fields on a sphere was carried out by \cite{1987MNRAS.226..655B}, but we follow notation of \cite{1998MNRAS.297..355S} used to characterize Minkowski functional estimators (which is similar to the task at hand). The covariance matrix $\Sigma$ can be written as 
\begin{equation}
  \Sigma = \left[\begin{matrix}
    \sigma & 0 & 0 & -\tau & -\tau & 0\\
    0 & \tau & 0 & 0 & 0 & 0\\
    0 & 0 & \tau & 0 & 0 & 0\\
    -\tau & 0 & 0 & v & \frac{v}{3} & 0\\
    -\tau & 0 & 0 & \frac{v}{3} & v & 0\\
    0 & 0 & 0 & 0 & 0 & \frac{v}{3}\\
  \end{matrix}\right],
\end{equation}
where $\sigma$, $\tau$, and $v$ are characteristic parameters of the Gaussian random field, which are derived from its correlation function $\xi(\theta)$ as \cite{Tomita:1990}
\begin{equation}\label{A.5}
  \sigma = \xi(0), \hspace{1em}
  \tau = |\xi''(0)|, \hspace{1em}
  v = \xi''''(0) .
\end{equation}
Prefactor multiplying the exponential function in Equation~\eqref{A.3} ensures the normalization condition
\begin{equation}
  \int P(\vec{n})\, {d\;\!}^6n = 1 .
\end{equation}
The parameters defined in Equation \eqref{A.5} determine the variances of the smoothed signal map $\tilde{s}$ and its derivatives, and thus of the morphological features that are contained in the vector ${\cal F}$. For small $\vec{n}$, these can be computed analytically using perturbation theory as follows. Let $X$ be an element in the feature space ${{\cal F}^{(n)}}$. The corresponding average $\ev{X}$ can be computed using the probability density function from Eq. \eqref{A.3} in the usual way
\begin{equation}
  \ev{X} = \int X(\vec{n})\, P(\vec{n})\, {d\;\!}^6n .
\end{equation}
The covariance between two elements $X$ and $Y$ in the feature space can be computed as
\begin{equation}
    \cov{X,Y}=\ev{(X-\ev{X})(Y-\ev{Y})} ,
\end{equation}
and the variance of a single element $X$ can be expressed as
\begin{equation}
    \var{X}=\cov{X,X} .
\end{equation}
After expanding non-linear expressions for feature vector components (\ref{eq:feature:12}, \ref{eq:feature:3}) to linear order in $\vec{n}$, and averaging over Gaussian noise ensemble (\ref{A.3}), the covariance matrix of the elements in our chosen feature space can be straightforwardly calculated, albeit after some lengthy algebra. Using the notation introduced earlier and dropping tilde on the source field $\tilde{s}$, the expectation values of the three features can be expressed as
\begin{equation}
\begin{aligned}
&\bar{\cal F}^{(1)} = s,\\
&\bar{\cal F}^{(2)} = |\nabla s|,\\
&\bar{\cal F}^{(3)} = \frac{(s_{11}-s_{22}) s_1 s_2 - s_{12}(s_1^2 - s_2^2)}{s_1^2 + s_2^2}.
\end{aligned}
\end{equation}
The corresponding covariance matrix is diagonal, i.e.\ the cross-covariance components $\cov{\delta{\cal F}^{(i)},\delta{\cal F}^{(j)}}$ with $i\ne j$ vanish, while diagonal elements are
\begin{equation}
\begin{aligned}
&\var{\delta{\cal F}^{(1)}} = \sigma,\\
&\var{\delta{\cal F}^{(2)}} = \tau,\\
&\var{\delta{\cal F}^{(3)}} = \frac{v}{3} + \rho\tau,
\end{aligned}
\end{equation}
where
\begin{equation} \label{eq:variance:rho}
  \rho = \frac{\left( (s_{11}-s_{22})(s_1^2 - s_2^2) + 4s_{12} s_1 s_2 \right)^2}{(s_1^2 + s_2^2)^3} .
\end{equation}
It is amusing to note that the numerator in the last expression corresponds to the expression in third Minkowski functional ${\cal I}_2$ with the trace of the Hessian times the gradient $(\nabla s)^2$ added. In practice, smoothed estimate $\tilde{s}$ is used instead of the true signal, and position-dependent expression (\ref{eq:variance:rho}) could be replaced by its average over the map $\bar{\rho}$ without too much ill effects. Thus, the covariance matrix $\cov{\delta{\cal F}}$ of the perturbed feature vector ${\cal F}$ can be taken to be
\begin{equation}
  \cov{\delta{\cal F}} = \left[\begin{matrix}
    \sigma & 0 & 0\\
    0 & \tau & 0\\
    0 & 0 & \frac{v}{3} + \bar{\rho}\tau
  \end{matrix}\right] .
\end{equation}
It is proportional to the noise variance $\var{\delta\tilde{s}}$, and could be scaled with the noise amount, separating the spectrum-dependent correlations from the overall amplitude.

\section{Gaussian noise correlation function}
\label{sec:app:correlation}

A Gaussian random field can be fully characterized by its angular two-point correlation function, related to its spectrum by
\begin{equation}\label{B.1}
  \xi(\theta) = \frac{1}{4\pi}\sum_{\ell=1}^{\infty} (2l+1)\, C_{\ell}\, P_{\ell}(\cos\theta) .
\end{equation}
This function describes the correlation between two points on a sphere as a function of their angular separation $\theta$. It is expressed as a sum over all values of the angular power spectrum $C_{\ell}$, weighted by the Legendre polynomials $P_\ell$ of degree $\ell$. The constant term ($\ell=0$) is typically removed from the sum. The characteristic parameters associated with Gaussian random field, as defined in Eq. \eqref{A.5}, can be computed as follows
\begin{equation}\label{B.2}
\begin{aligned}
&\xi(0) = \frac{1}{4\pi}\sum_{\ell=1}^{\infty} (2\ell+1)\, C_{\ell},\\
&\xi''(0) = -\frac{1}{4\pi}\sum_{\ell=1}^{\infty} (2\ell+1)\, \frac{\ell(\ell+1)}{2}\, C_{\ell},\\
&\xi''''(0) = \frac{1}{4\pi}\sum_{\ell=1}^{\infty} (2\ell+1)\, \frac{\ell(\ell+1)(3\ell^2+3\ell-2)}{8}\, C_{\ell}.
\end{aligned}
\end{equation}
These expressions typically diverge for high $\ell$ for white or even scale-invariant noise spectra, but they are naturally cut-off by finite resolution of the experiment, which could be described by an effective window function \citep{White:1994sx}. For a Gaussian cut-off, the underlying angular power spectra $C_{\ell}$ are multiplied by a factor of $B_{\ell}^{2}$, where the beam transfer function is given by
\begin{equation}
  B_{\ell}=\exp\left[-\frac{1}{2}\ell(\ell+1)\, \delta^2\right] .
\end{equation}
The parameter $\delta$ is related to the FWHM of the beam $\theta_{\textrm{FWHM}}$ by
\begin{equation}
    \delta^2 = \frac{\theta_{\textrm{FWHM}}^2}{8\ln2} .
\end{equation}
This convolution introduces a high-$\ell$ cutoff at scales roughly inversely corresponding to the beam size $\ell\sim1/\delta$. For more detailed exposition, see \cite{1995ApJ...443....6W}.

The power spectrum of white Gaussian noise, accounting for the effects of finite beam resolution, is thus given by
\begin{equation}
    C_\ell \propto \exp\left[-\ell(\ell+1)\, \delta^2\right] .
\end{equation}
To obtain analytic expressions for the characteristic parameters of the Gaussian random field \eqref{A.5}, we can approximate the sum in \eqref{B.2} with a continuum integral
\begin{equation}
\sum_{\ell=1}^{\ell_{\text{max}}} \to \int\limits_{0}^{\ell_{\text{max}}}d{\ell} ,
\end{equation}
which is valid for $l_{\text{max}}\gg 1$. The resulting expressions are then
\begin{equation}
\begin{aligned}
&\sigma = \frac{\delta^{-2}}{4\pi}, \\
&\frac{\tau}{\sigma} =  \frac{\delta^{-2}}{2}, \\
&\frac{v}{\sigma} = \frac{\delta^{-4}}{4}\, (3-\delta^2).
\end{aligned}
\end{equation}

For a scale-invariant noise model, the measured power spectrum considering finite beam resolution is given by 
\begin{equation}
  C_{\ell} \propto \frac{1}{\ell(\ell+1)}\, \exp\left[-\ell(\ell+1)\, \delta^2\right] .
\end{equation}
To calculate the characteristic parameters \eqref{A.5} in this case, we can use the following approximation for discrete sum
\begin{equation}
  \sum\limits_{\ell=1}^{\ell_{\text{max}}} \frac{1}{\ell(\ell+1)} \to
  \int\limits_{\frac{1}{2}}^{\ell_{\text{max}}} \frac{d\ell}{(\ell+\frac{1}{2})^2} ,
\end{equation}
which also holds for large $l_{\text{max}}$. The parameters can then be expressed in terms of the exponential integral function $E_{1}$ as
\begin{equation}
\begin{aligned}
&\sigma = \frac{\textrm{E}_{1}(\delta^2)}{4\pi}\, e^{\delta^2/4}, \\
&\frac{\tau}{\sigma} = \frac{1}{2}\frac{\delta^{-2}}{\textrm{E}_{1}(\delta^2)\, e^{\delta^2}} - \frac{1}{8}, \\
&\frac{v}{\sigma} = \frac{\delta^{-4}}{16}\frac{6 - \delta^{2}}{\textrm{E}_{1}(\delta^2)\, e^{\delta^2}} + \frac{11}{128}.
\end{aligned}
\end{equation}
These expressions can be numerically evaluated using the series expansion
\begin{equation}
  \textrm{E}_{1}(\delta^2) = -\gamma - 2\ln\delta + O(\delta^2),
\end{equation}
where $\gamma$ is the Euler-Mascheroni constant.

\section{Preprocessing of 353GHz dust map}
\label{sec:app:dust}

As an example of strongly non-Gaussian signal contaminated with non-uniform pointwise Gaussian noise, we use dust maps derived from 353GHz Planck 2018 data \citep{2020A&A...641A...3P}. While more sophisticated treatment might be in order for optimal component separation \citep{2008A&A...491..597L, 2014A&A...571A..11P, 2020A&A...641A...4P}, we opt for the minimally processed option for simplicity. In this approach, CMB and point source components are subtracted from 353GHz map, otherwise dominated by (polarized) dust emission, without any leakage or bandpass corrections. Complete details on the processing are as follows.

First, the effective beam function of 353GHz maps (\href{https://wiki.cosmos.esa.int/planck-legacy-archive/index.php/Effective_Beams}{available with Planck 2018 data}) is matched to component-separated CMB maps published by Planck collaboration (which are provided with effective $5'$ FWHM Gaussian beam at $\nside = 2048$ resolution). This is accomplished by full-sky convolution using fast spherical transform. Next, the CMB component is subtracted out. We use SMICA component-separated map for that. Then, estimated cosmic infrared background monopole
is subtracted out, and galactic offset
is added \citep{2014A&A...571A..11P}. This has no bearing on noise reduction algorithms discussed in this paper, but is quite important if one wants to estimate polarization fraction of the dust radiation.

Next, contribution of resolved point sources is estimated and subtracted out from intensity map (point sources are typically not highly polarized). Identification of point sources is carried out by applying isotropic Wiener filter to 353GHz, 545GHz, and 847GHz maps (matching best-fit power-law of dust emission at a given frequency to white spectrum asymptotic of point sources), and identifying point sources as regions above $3\sigma$ in Wiener-filtered maps (using robust estimator for $\sigma$ that rejects roughly $20$\% of the tails), and connected parts determined by watershed algorithm. To ensure smooth transitions, the thresholded point source map is inpainted outside the source mask by solving massive scalar field equation $\Delta\varphi = m^2\varphi$ using multi-grid solver \citep{Brandt:1977, 2020A&A...641A...7P}. This assures smooth fall-off away from the sources at a finite distance, just like in Yukawa potential. 

Once the resolved point sources are subtracted from intensity, the resultant map is a fairly good approximation of dust emission. There is still some measurable contamination from unresolved point sources evident in the anisotropy spectra, as well as a certain amount of sub-dominant foregrounds present, but on the whole the map is pretty good. Certainly it meets the requirements of a test sample for noise cleanup, which is the main subject of this paper.

\end{document}